\documentclass[aps,prl,reprint,showpacs,amsmath,amssymb,floatfix,superscriptaddress,noeprint]{revtex4-1}

\usepackage{placeins}
\usepackage[colorlinks,linkcolor=blue,urlcolor=blue,citecolor=blue,anchorcolor=blue]{hyperref}

\usepackage{graphicx} 
\usepackage{lipsum}

\usepackage{physics}
\usepackage{siunitx}
\DeclareSIUnit{\belmilliwatt}{Bm}
\DeclareSIUnit{\dBm}{\deci\belmilliwatt}
\usepackage{tensor}

\usepackage{xcolor}

\usepackage{stackengine}
\stackMath

\newcommand{\rhohat}{\ensuremath{\hat{\rho}}}

\newcommand{\ahat}{\ensuremath{\hat{a}}}
\newcommand{\adag}{\ensuremath{\hat{a}^\dagger}}

\newcommand{\cdag}{\ensuremath{\hat{c}^\dagger}}
\newcommand{\chat}{\ensuremath{\hat{c}}}

\newcommand{\sigy}{\ensuremath{\hat{\sigma}_y}}
\newcommand{\sigx}{\ensuremath{\hat{\sigma}_x}}

\newcommand{\sigm}{\ensuremath{\hat{\sigma}_-}}

\makeatletter
\newsavebox\myboxA
\newsavebox\myboxB
\newlength\mylenA

\newcommand*\xoverline[2][0.9]{%
    \sbox{\myboxA}{$\m@th#2$}%
    \setbox\myboxB\null
    \ht\myboxB=\ht\myboxA%
    \dp\myboxB=\dp\myboxA%
    \wd\myboxB=#1\wd\myboxA
    \sbox\myboxB{$\m@th\overline{\copy\myboxB}$}
    \setlength\mylenA{\the\wd\myboxA}
    \addtolength\mylenA{-\the\wd\myboxB}%
    \ifdim\wd\myboxB<\wd\myboxA%
       \rlap{\hskip 0.5\mylenA\usebox\myboxB}{\usebox\myboxA}%
    \else
        \hskip -0.5\mylenA\rlap{\usebox\myboxA}{\hskip 0.5\mylenA\usebox\myboxB}%
    \fi}
\makeatother

\begin{document}
\renewcommand\stackalignment{l}
\title{Monitoring the energy of a cavity by observing \\the emission of a repeatedly excited qubit}

\author{H. Hutin}
\affiliation{Ecole Normale Sup\'erieure de Lyon,  CNRS, Laboratoire de Physique, F-69342 Lyon, France}
\author{A. Essig}
\affiliation{Ecole Normale Sup\'erieure de Lyon,  CNRS, Laboratoire de Physique, F-69342 Lyon, France}
\author{R. Assouly}
\affiliation{Ecole Normale Sup\'erieure de Lyon,  CNRS, Laboratoire de Physique, F-69342 Lyon, France}
\author{P. Rouchon}
\affiliation{Laboratoire de Physique de l'Ecole normale sup\'erieure,  Mines-Paris - PSL, Inria, ENS-PSL, CNRS,  Universit\'e PSL, Paris, France. }
\author{A. Bienfait}
\affiliation{Ecole Normale Sup\'erieure de Lyon,  CNRS, Laboratoire de Physique, F-69342 Lyon, France}
\author{B. Huard}
\affiliation{Ecole Normale Sup\'erieure de Lyon,  CNRS, Laboratoire de Physique, F-69342 Lyon, France}

\date{\today}
\begin{abstract}
The number of excitations in a large quantum system (harmonic oscillator or qudit) can be measured in a quantum non demolition manner using a dispersively coupled qubit. It typically requires a series of qubit pulses that encode various binary questions about the photon number. Recently, a method based on the fluorescence measurement of a qubit driven by a train of identical pulses was introduced to track the photon number in a cavity, hence simplifying its monitoring and raising interesting questions about the measurement backaction of this scheme. A first realization with superconducting circuits demonstrated how the average number of photons could be measured in this way. Here we present an experiment that reaches single shot photocounting and number tracking owing to a cavity decay rate 4 orders of magnitude smaller than both the dispersive coupling rate and the qubit emission rate. An innovative notch filter and pogo-pin based galvanic contact makes possible these seemingly incompatible features. The qubit dynamics under the pulse train is characterized. We observe quantum jumps by monitoring the photon number via the qubit fluorescence as photons leave the cavity one at a time. Besides, we extract the measurement rate and induced dephasing rate and compare them to theoretical models. Our method could be applied to quantum error correction protocols on bosonic codes or qudits.
\end{abstract}
\maketitle

\begin{figure}[!t]
\centering
\includegraphics[width=\columnwidth]{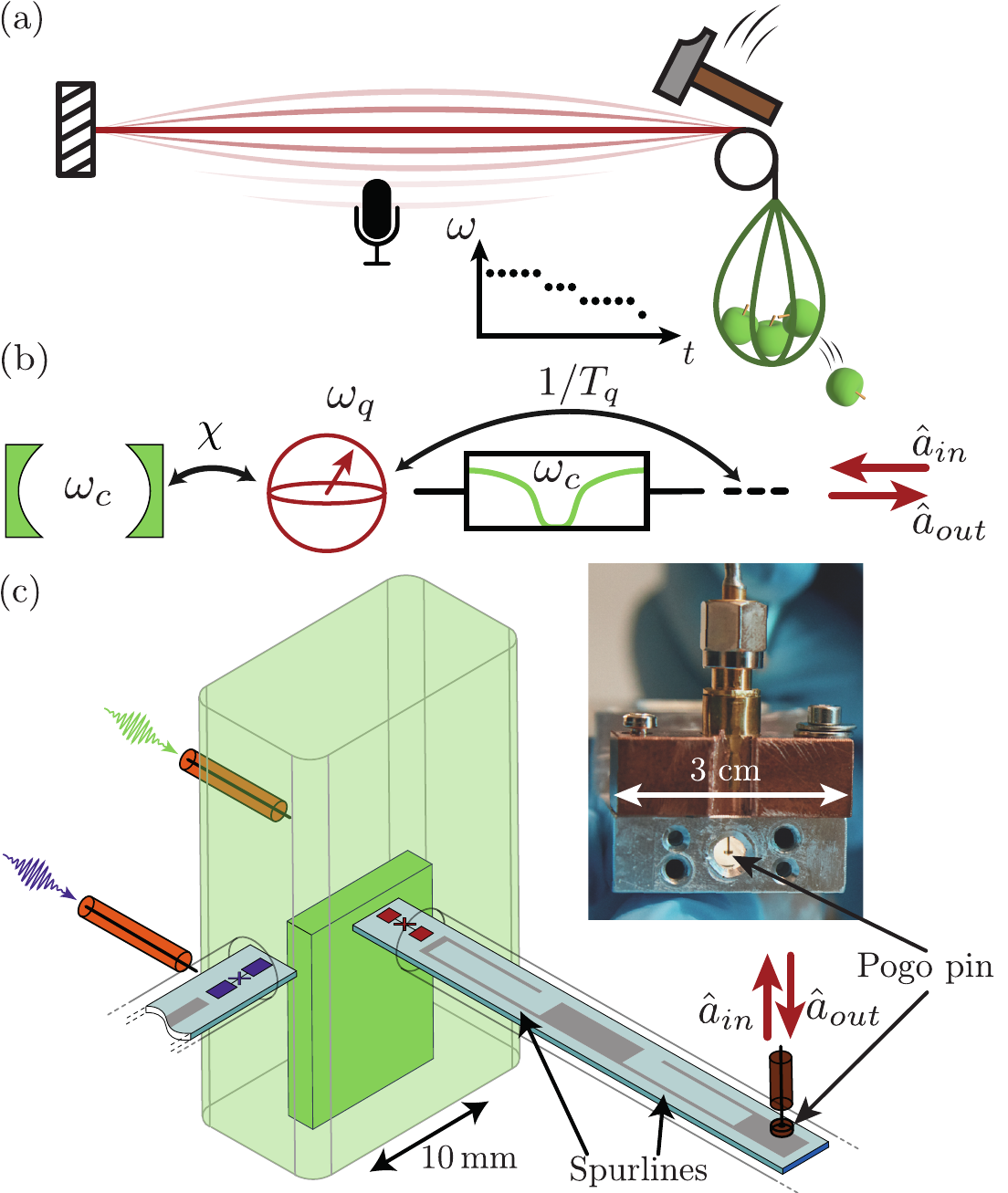}
\caption{(a) Acoustic analogy of the experiment. In order to monitor the number of apples in a basket, it is attached to a string. By repeatedly hitting the string, recording the emitted sound reveals the number $n$ at any time. (b) The heterodyne detection of the driven qubit (red) emission into the line leads to the monitoring of the photon number of the dispersively coupled cavity (green). A notch filter centered on $\omega_c$ (box) prevents the cavity from decaying through the qubit.
(c) The cavity is a high purity aluminum $\lambda/4$ coaxial resonator. Two transmon qubits
on sapphire chips stick into the cavity. The filter on the (red) qubit drive line is composed of two tantalum spurlines and galvanically connected to a transmission line using a pogo pin (photograph in inset). Wigner tomography of the cavity is performed using the auxiliary transmon qubit (purple) and its dedicated readout resonator (not shown).}
\label{fig:figure_1}
\end{figure}

Resolving the number of photons in an electromagnetic mode is at the core of many quantum information protocols~\cite{Knill2001,Simon2007,Aaronson2014,Hamilton2017}. Most of them require quantum nondemolition (QND) measurements for measurement based feedback or heralding. In the microwave domain, such measurements can be performed using dispersively coupled Rydberg atoms~\cite{Gleyzes2007} or superconducting qubits~\cite{Schuster2007}. Predetermined~\cite{Guerlin2007,Johnson2010} or adaptive~\cite{Peaudecerf2013,Peaudecerf2014,Wang2020,Dassonneville2020,Curtis2020} measurement sequences can monitor the photon number in time and even detect quantum jumps. Each measurement step yields at most a single bit of information about the photon number. By adapting each step, it is possible to reach this upper bound and photocount in a number of cycles that scales logarithmically with the maximal photon number~\cite{Wang2020,Dassonneville2020,Curtis2020}. 
Recently, another qubit-based detector was introduced, which is able to track the photon number using a train of identical qubit pulses forming a frequency comb~\cite{Essig2021}. Consequently the frequency of the qubit fluorescence encodes the photon number at any time. While a proof-of-principle experiment demonstrated signals proportional to the photon number~\cite{Essig2021}, the measurement rate was insufficient compared to the cavity lifetime for single-shot extraction. Here, we demonstrate photon number tracking in a 3D cavity using a frequency comb driving a dispersively coupled qubit. 
We experimentally compare the photon number measurement rate of our scheme based on heterodyne detection of the qubit fluorescence to the rate at which the environment could extract information. 
This QND photon number monitor, with a fixed drive and detection scheme, could simplify feedback schemes and quantum error correction for bosonic codes~\cite{Cai2021} or qudits~\cite{Joo2019,Wang2020a}.

The detection principle can easily be grasped with a classical analogy (Fig.~\ref{fig:figure_1}a). Consider a basket (cavity) filled with apples (photons) that can escape. The number $n$ of apples can be determined by hitting a string (qubit) from which the basket hangs. The string oscillates at a frequency depending on $n$ and recording the emitted sound (heterodyne measurement of fluorescence signal) reveals the apple number. Hitting repeatedly the string leads to the monitoring of $n$: in the frequency domain, it corresponds to driving the string with a comb. 

Experimentally, the basket is an aluminum $\lambda/4$ coaxial cavity~\cite{Reagor2015a} at frequency $\omega_c/2\pi = 4.573~\mathrm{GHz}$.  The qubit is a transmon at $\omega_q/2\pi = 6.181~\mathrm{GHz}$, dispersively coupled to the cavity with a frequency shift $-\chi/2\pi = -5.25$ MHz per photon. Photon number tracking requires to operate in the photon number resolved regime $\chi>\Gamma_2$, with $\Gamma_2/2\pi\approx 3.5~\mathrm{MHz}$ the qubit coherence rate~\cite{Essig2021}. To optimize the information rate, we maximize the qubit emission rate $1/T_q\leq 2\Gamma_2$ into the measurement line under this constraint. 

To protect the cavity from decaying through the qubit, we use a notch filter at the cavity frequency (Fig.~\ref{fig:figure_1}b). 
The filter circuit is composed of two on-chip spurlines in series (Fig.~\ref{fig:figure_1}c). The qubit and the notch filter are patterned out of a tantalum film on the same sapphire chip, which is inserted into the cavity~\cite{supmat}. A galvanic connection is ensured between the measurement line and the on-chip filter using an SMA microwave connector terminated by a pogo pin~\footnote{POGO-PIN-19.0-1 by Emulation Technology}, which is a pin connected to a spring (Fig.~\ref{fig:figure_1}c). This filter design leads to a high coupling rate $1/T_{q}=(23\pm 3~\mathrm{ns})^{-1}$ between the qubit and the measurement line, while preserving a cavity lifetime $T_c$ larger than $200~\mu\mathrm{s}$ for a single photon. The heterodyne detection benefits from the large bandwidth of  a traveling wave parametric amplifier (TWPA~\cite{Macklin2015}) that covers many $\chi$.  An auxiliary transmon qubit and its readout resonator are used to perform direct Wigner tomography of the cavity state~\cite{Lutterbach1997,Bertet2002,Vlastakis2013,supmat}.

\begin{figure}[!ht]
\centering
\includegraphics[width=\linewidth]{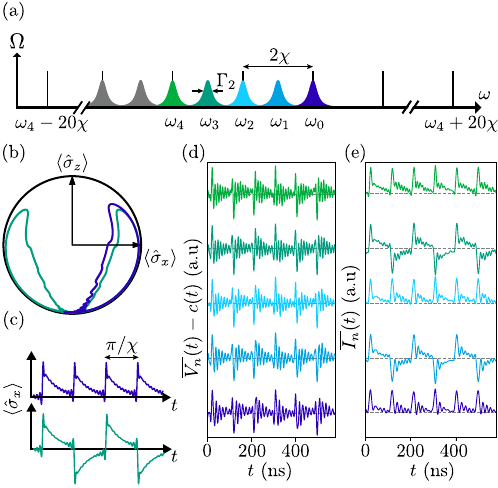}
\caption{(a) Black peaks: driving comb amplitude $\Omega$ Eq.~(\ref{eq:1}) in the frequency domain. Colored Lorentzian shapes:  qubit frequency spectra corresponding to cavity photon numbers from 0 to 6. (b) Simulated average trajectories of the qubit in the Bloch sphere for $n = 0$ (blue) and $n = 1$ (green). The kick angle is set to $\theta = \pi\Omega/\chi=3\pi/4$, and as in the experiment, $\chi T_q=0.76$. (c) Corresponding average values of $\langle\sigx\rangle$ as a function of time. 
(d) Averaged measured record $\overline{V_n}(t)$ when the cavity is prepared in a Fock state $|n\rangle$ from $n=0$ to 4 from bottom to top, and for a comb amplitude such that $\theta = \pi\Omega/\chi\approx \pi/2$. The curves are offset by the much larger driving comb signal $c(t)$. (e) Corresponding quadrature $\overline{I_n}(t)$ of the emitted fluorescence signal reconstructed (see text) at each qubit frequency $\omega_q-n\chi$.} 
\label{fig:figure_2}
\end{figure}

The qubit is driven with a frequency comb of amplitude $\Omega$ and peaks at $\omega_q+k\Delta\omega$ where $k$ spans all integers between $-K$ and $K$. In the lab frame, the qubit drive Hamiltonian thus reads 
\begin{equation}\label{eq:1}
\hat{H}_{d} = -\hbar \frac{\Omega}{2} \sum_{k = -K}^{K} \cos \left[(\omega_q+ k \Delta\omega) t\right]\sigy.
\end{equation}
In the limit of an infinite Dirac comb ($K\rightarrow\infty$), it becomes a series of Dirac peaks in the time domain with a period $2\pi/\Delta\omega$. In the frame rotating at $\omega_q-\chi\cdag\chat$, and under the rotating wave approximation, it gives
\begin{equation}
\hat{H}_{d} = \hbar \frac{\pi\Omega}{\Delta\omega} \sum_{l = -\infty}^{\infty} \delta\left(t- \frac{2\pi l}{\Delta\omega}\right)\hat{\sigma}(t),\label{eq:2}
\end{equation}
where $\hat{\sigma}(t)=\sin(\cdag \chat \chi t)\sigx-\cos (\cdag \chat \chi t)\sigy$ and $\cdag\chat$ is the photon number in the cavity.
The dynamics of the qubit Bloch vector thus consists in periodic kicks every $2\pi/\Delta\omega$ by an angle $\theta=2\pi\Omega/\Delta\omega$. The natural choice is to have one peak per possible qubit frequency $(\Delta\omega=\chi)$. The rotation axis of the kicks would then be the same for any photon number since $\varphi=2\pi\cdag\chat\chi/\Delta\omega$ is a multiple of $2\pi$. 
However, the period between two kicks would then be $2\pi/\chi\approx 190~\mathrm{ns}$, which is much longer than  $T_q$. To limit idle times in the qubit fluorescence signal, we choose a twice larger peak spacing $\Delta\omega=2\chi$, which doubles the information rate
(Fig.~\ref{fig:figure_2}a).  Consequently, $\varphi$ is equal to $0$ mod $2\pi$ for even photon numbers and $\pi$ mod $2\pi$ for odd photon numbers. Therefore the kick direction flips with each kick for odd photon numbers.

In the experiment, we choose a finite number $2K+1=21$ peaks in the comb. In the frequency domain the drive is the product of a square window of width $21\chi/2\pi$ and the infinite comb. Consequently, in the time domain, the resulting waveform is the convolution of the infinite comb with a sinc function.
This width sets the timescale of each qubit kick to $\pi/21\chi\approx 5~\mathrm{ns}$, which is much shorter than $T_q$. Additionally, this choice guarantees that the qubit frequency remains well within the bandwidth of the comb, regardless of the desired photon number ranging from 0 to $N_\mathrm{max}=9$. To further minimize boundary effects, we position the comb center frequency at $\omega_q-4\chi$, which corresponds to the qubit frequency associated with 4 photons in the cavity.

The average predicted dynamics of the qubit is shown in Fig.~\ref{fig:figure_2}b,c when the cavity has 0 (blue) or 1 (green) photon. At intervals of $\pi/\chi$, the qubit state undergoes a kick lasting approximately $5~\mathrm{ns}$, followed by relaxation as it fluoresces into the measurement line. In contrast with even photon numbers, the rotation axis flips at every kick for odd photon numbers. Heterodyne detection of the fluorescence field measures the two quadratures $(\adag_\mathrm{out}+\ahat_\mathrm{out})/2$ and $i(\ahat_\mathrm{out}-\adag_\mathrm{out})/2$. The emitted field amplitude can be expressed as $\ahat_\mathrm{out}=\ahat_\mathrm{in}-\hat{\sigma}_-/\sqrt{T_q}$, where $\langle\ahat_\mathrm{in}\rangle$ is the driving comb and $\sigm=(\sigx-i\sigy)/2$ is the qubit lowering operator~\cite{Gardiner1985}. The average dynamics of the qubit coherences (Fig.~\ref{fig:figure_2}c) can thus be directly observed in the heterodyne signal. 

We first prepare a Fock state $|n\rangle$ using a coherent excitation on the cavity followed by heralding using the qubit emission under a drive at a single tone $\omega_q-n\chi$~\cite{Gely2019,Essig2021,supmat}.  
We then apply the comb. The qubit fluorescence is amplified and downconverted by a local oscillator at $\omega_q+\omega_\mathrm{IF}$, with $\omega_\mathrm{IF}=2\pi\times 66~\mathrm{MHz}$. The amplified fluorescence signal is recorded as a voltage $V(t)$ using an analog-to-digital-converter. The average $\overline{V_n}(t)$ of these records under the heralding of $n$ photons is shown in Fig.~\ref{fig:figure_2}d offset by the contribution of the reflected driving comb $c(t)\propto\mathrm{Re}(\langle \hat{a}_\mathrm{in}\rangle)$~\cite{supmat}. These signals can be processed to reveal the evolution of  $\langle\sigx\rangle$ and $\langle\sigy\rangle$ in the qubit frame when there are $n$ photons. To do so, we extract their analytic representations~\cite{supmat} and demodulate them at $\omega_\mathrm{IF}+n\chi$ to obtain two average quadratures $\overline{I_n}(t)$ and $\overline{Q_n}(t)$. The traces of  $\overline{I_n}(t)$ are shown in Fig.~\ref{fig:figure_2}e for $n=0$ to $4$ and match the expected evolution of $\langle\sigx\rangle$. The kicks and the subsequent decays are visible. The kick direction alternates for odd numbers of photons as expected. The remaining oscillations in the reconstructed signal may be due to an imperfect subtraction of the driving comb $c(t)$, or to a distortion of the driving comb or output signal by the measurement setup. 

\begin{figure}[!h]
\centering
\includegraphics[width=\linewidth]{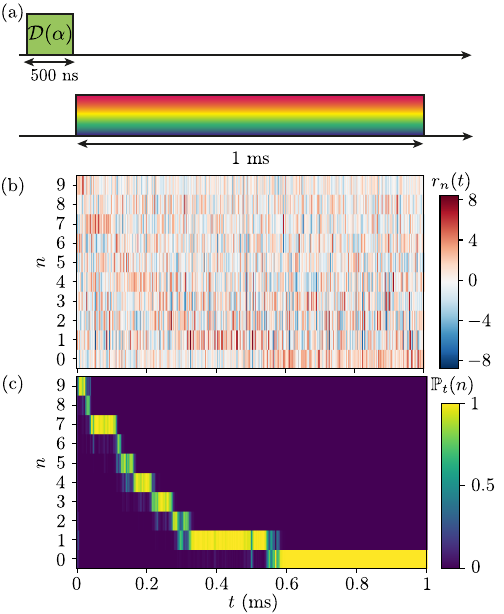}
\caption{(a) Pulse sequence for the observation of quantum jumps. (b) Evolution of the measurement outcomes $r_n$ as a function of time $t$. The measurement is performed on one realization using a driving comb whose amplitude corresponds to an expected kick rotation angle $\theta=\pi/2$. (c) Evolution of the inferred probability distribution $\mathbb{P}_t(n)$ of the photon number $n$ using the outcomes in (b). 1000 realizations in the same conditions are available in~\cite{hutin_2024_trajectories}.}
\label{fig:figure_3}
\end{figure}

Decoding the measurement record $V(t)$ in order to infer the photon number is a task similar to quantum sensing using continuous measurement~\cite{Gambetta2001,Gammelmark2013b,Gammelmark2014,Catana_2015,Kiilerich2016,Albarelli2018,Albarelli2019,Ilias2022,Descamps2022}. Here  we use the average records $\overline{V_n}$ as demodulation weight functions, and define $N_\mathrm{max}$+1 measurement outcomes represented as a vector $\vec{m}$ whose components are
\begin{equation}
m_n(t) = \int_{t-\tau}^{t} V(t') \overline{V_n}(t')\mathrm{d}t',
\end{equation}
where the integration time $\tau=2~\mu\mathrm{s}$ is chosen much shorter than the cavity lifetime $T_c$ and multiple $(21\times)$ of $\pi/\chi$. As a demonstration, we excite the cavity with a coherent state with more than 20 photons on average using a strong resonant pulse, then drive the qubit with the frequency comb and record $\vec{m}$ as a function of time $t$. Quantum jumps on a single realization can already be visualized by a simple data processing. We perform a time independent linear transform~\cite{supmat} $\vec{r}(t)=\mathbf{G}^{-1}\vec{m}(t)$ so that, on average, $r_n(t)=1$ for n photons in the cavity, while all the other components $r_{k\neq n}$ vanish. Concretely, $\mathbf{G}$ is the Gram matrix of the average records $\overline{V_n}(t)$ so that $\mathbf{G}_{nm}=\int_{0}^{\tau} \overline{V_n}(t) \overline{V_m}(t)\mathrm{d}t$. The evolution of $\vec{r}$ is shown in Fig.~\ref{fig:figure_3}b for one realization of the experiment. A faint red trace emerges from the noise, which reveals the successive losses of single photons in the cavity that here decays from 9 to 0 photons over 1~ms.

To predict the number $n$ of photons at any time of the evolution, the probability distribution of  $n$ is updated conditionally on the outcome $\vec{m}(t)$ through Bayesian update at every time step $j\tau$. It first requires determining the likelihood $\mathbb{P}(\vec{m}|n)$ conditioned on the cavity being in the Fock state $|n\rangle$. $\mathbb{P}(\vec{m}|n)$ can be approximated by a Gaussian function of $\vec{m}$ owing to the small measurement efficiency $\eta=0.17\pm 0.02$. Therefore, we characterize its distribution by the measured mean $\langle\vec{m}\rangle_{|n\rangle}$ and covariance matrix of $\vec{m}$ for each $|n\rangle$ only~\cite{supmat}. Using this procedure, along with accounting for photon loss during each time step $j\tau$, the noisy measurement outcomes $\vec{m}(j\tau)$ of Fig.~\ref{fig:figure_3}b lead to the probability distribution $\mathbb{P}_{j\tau}(n)$ shown in Fig.~\ref{fig:figure_3}c for the same realization. Note that we assume no prior information ($\mathbb{P}_0(n)=1/10$), but this choice has anyway no impact on the quantum trajectory after a few $\tau$. 
With many realizations, we extract the average photon number decay. Interestingly, it is not exponential, which indicates a subtle interplay between cavity dissipation and qubit dynamics~\cite{supmat}.
\begin{figure}
\centering
\includegraphics[width=\linewidth]{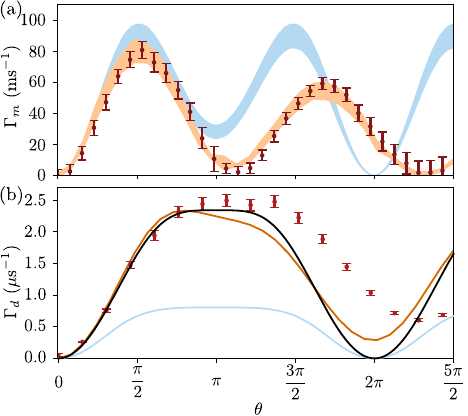}
\caption{(a) Red dots: observed average measurement rate $\Gamma_m$ as a function of drive amplitude, expressed as the qubit expected rotation angle $\theta$ per kick. 
Orange shadow: measurement rate obtained using a stochastic master equation with detection efficiency $\eta$ spanning the range $[0.17,0.20]$. Blue shadow: theoretical bound for an infinite comb and heterodyne measurement with efficiencies $\eta \in [0.17, 0.20]$.
(b) Red dots: observed cavity dephasing rate $\Gamma_d$ as a function of $\theta$. Orange: simulated measurement induced dephasing rate. Black: theoretical accessible information rate. 
Blue: theoretical maximal ($\eta = 1$) measurement rate obtained by heterodyne detection. }
\label{fig:figure_4}
\end{figure}

We now determine the measurement rate $\Gamma_m$ of the photon number.  Formally, it is the time derivative of the mutual information between the photon number $n$ and the outcome $\vec{m}$ at $t=0$~\cite{Clerk2003,Clerk2008}. In the weak measurement regime $\tau\Gamma_m\ll 1$, it can be approximated by
\begin{equation}\label{MI_hetero}
\begin{split}
\tau\Gamma_m  = &-\int \mathbb{P}(\vec{m})\log\mathbb{P}(\vec{m})\mathrm{d}\vec{m}\\
&+\sum_n \mathbb{P}_0(n)\int \mathbb{P}(\vec{m}|n)\log\mathbb{P}(\vec{m}|n)\mathrm{d}\vec{m},
\end{split}
\end{equation}
where $\mathbb{P}(\vec{m})=\sum_n \mathbb{P}_0(n) \mathbb{P}(\vec{m}|n)$. As shown in Fig.~\ref{fig:figure_3}c, at most two photon numbers are likely at any time. We thus choose the prior $\mathbb{P}_0(q)=\mathbb{P}_0(q+1)=1/2$, and average over all $q$ values 
to compute a measurement rate $\Gamma_m$~\cite{supmat}. Plugging in Eq.~(\ref{MI_hetero}) the measured distributions $\mathbb{P}(\vec{m}|n)$ at various driving amplitudes $\Omega$, $\Gamma_m$ is obtained as a function of the kick angle $\theta=\pi\Omega/\chi$ (Fig.~\ref{fig:figure_4}a). It is maximal when the qubit is kicked to states corresponding to the largest $\langle\sigx\rangle$. It can be intuitively understood since heterodyne detection probes the quadratures of the emitted fluorescence signal, which are on average proportional to the coherence $\langle\sigx-i\sigy\rangle$. Notably, the rate $\Gamma_m$ is much larger than the cavity decay rate $1/T_c$ (more than 16 times), which is well in the single shot measurement regime. On average, this measurement scheme allows us to reach more than 95\% average confidence in 20 $\mu$s ~\cite{supmat}. Our complete model~\cite{supmat} reproduces the observed rate $\Gamma_m$ using simulated measurement records by a stochastic master equation with detection efficiency $\eta=0.185\pm 0.015$ as a fit parameter, in agreement with the independently measured $\eta$.

It is interesting to compare this measurement rate to the rate at which information about the photon number leaks into the environment, \emph{i.e.} the cavity dephasing rate $\Gamma_d$. We compute it as the added decay rate of $\Tr{\rhohat(t)\ahat}$ on top of the natural decoherence rate~\cite{supmat}. 

We use the auxiliary transmon qubit (purple in Fig.~\ref{fig:figure_1}c) to perform Wigner tomography on the cavity state and extract $\Tr{\rhohat(t)\ahat} = \int \alpha W_t(\alpha)\mathrm{d}^2\alpha$, with $W_t(\alpha)$ the Wigner function of $\rhohat(t)$. The dependence of $\Gamma_d$ on kick angle is shown as red dots in Fig.~\ref{fig:figure_4}b. Strikingly, its maximum is reached at $\theta\approx\pi$, where the qubit has the largest energy to emit, and thus the most information to leak out. We note that the model that successfully predicts the measurement rate (orange) underestimates the measurement induced dephasing rate at large drive amplitudes, indicating that the driving comb leads to stronger decoherence than anticipated.

The dephasing rate $\Gamma_d$ is about 20 times larger than $\Gamma_m$ at $\theta=\pi/2$ (its maximum). Indeed our measurement setup does not recover the full information available because of a limited detection efficiency and the very use of heterodyne detection. To better understand this information loss, we use a simplified model where the comb is infinite. It reproduces the measured $\Gamma_m$ with $0.17<\eta<0.20$ (blue in Fig.~\ref{fig:figure_4}a) for small angles $\theta$, but not at larger angles where the finite duration of the kicks reduces the actual rotation angle.  However, the dephasing rate is larger than the measurement rate of an ideal ($\eta=1$) heterodyne measurement (blue in Fig.~\ref{fig:figure_4}b). Accordingly, with the same model, the upper bound on the measurement rate for any detection scheme -- accessible information rate -- (black in Fig.~\ref{fig:figure_4}b) is close to the measured $\Gamma_d$ for small angles $\theta$ and up to about 3 times larger than what the best heterodyne detector could do: Even an ideal heterodyne detector would destroy up to about 2/3 of the accessible information. The experiment provides here a textbook example of destroyed information by a measurement apparatus, here the heterodyne detector~\cite{Han2018}. 
In contrast to low detection efficiency, heterodyne measurement with $\eta=1$ would reveal information even for $\theta=\pi$. Indeed, while the average heterodyne signal is zero, its cumulants reveal the photon number. The signal-to-noise ratio on the cumulants of order $l$ scales as $\eta^l$ so that, in the experiment, the average is the main source of information, hence the minimum at $\theta=\pi$.

In conclusion, our superconducting circuit and signal processing demonstrate the possibility to monitor photon numbers in a cavity with a fixed driving. Information about the photon number is extracted up to 16 times faster than the cavity decay rate. The detector requires $20~\mu\mathrm{s}$ of measurement on average to reach 95~\% fidelity between 0 and 9 photons. A circuit using a harmonic oscillator or qudit instead of the qubit as an encoder could enable even faster measurement rate in future devices. A simple model quantitatively explains the dephasing and measurement rates for small drive amplitudes. As we look ahead to more integrated amplifiers, it would be interesting to observe how the driving amplitude that maximizes the measurement rate evolves with increased efficiency. Additionally, the pogo-pin and spurline filters offer a convenient architecture for achieving galvanic coupling with quantum circuits within a long lived microwave cavity. 
Interesting open questions remain to be explored such as the origin of the dependence of Fock state decay rates on the comb amplitude. This seems to be a dual effect to the readout problem of superconducting qubit~\cite{Sank2016,lescanne2018dynamics,Petrescu2020,Shillito2022,Khezri2022,Burgelman2022a,Thorbeck2023,Bengtsson2023}.

\FloatBarrier

\begin{acknowledgments}
This research was supported by the QuantERA grant ARTEMIS, by ANR under the grant ANR-22-QUA1-0004, and PR received funding from the European Research Council (ERC) under the European Union’s Horizon 2020 research and innovation programme (grant agreement No. [884762]). We acknowledge IARPA and Lincoln Labs for providing a Josephson Traveling-Wave Parametric Amplifier. We thank R\'emy Dassonneville, Pierre Guilmin, S\'ebastien Jezouin, Perola Milman, Mazyar Mirrahimi, Klaus M\o lmer, Alain Sarlette, Antoine Tilloy, Beno\^it Vermersch and Mattia Walschaers for useful discussions.
\end{acknowledgments}

\end{document}


\renewcommand\stackalignment{l}
\title{Supplementary materials: Monitoring the energy of a cavity by observing the emission of a repeatedly excited qubit}

\maketitle
\section{Device and measurement setup}
\subsection{Device fabrication}
The circuit is composed of one 3D $\lambda/4$ coaxial cavity resonator, into which two samples are inserted. The first sample comprises the monitoring transmon and its notch filter, the second contains the tomography transmon qubit with its readout resonator and Purcell filter. These samples are made of a 200 nm thick film of  sputtered Tantalum on a 430 $\mathrm{\mu m}$ thick sapphire substrate (deposited by Star Cryoelectronics, Santa Fe, USA). The Josephson junctions of both transmons are standard Dolan bridge e-beam evaporated Al/AlOx/Al junctions~\cite{thesisessig2021}.

\begin{table}\caption{Table of circuit parameters}
\begin{tabular}{ |p{6cm}||p{3cm}|p{3cm}|p{3cm}|  } 

 \hline
 \multicolumn{4}{|c|}{Table of circuit parameters} \\
 \hline
  \hline
 Circuit parameter& Symbol & Hamiltonian term & Value \\
 \hline
 Cavity frequency   & $\omega_c/2\pi$   &$\hbar \omega_c\adag\ahat$ &   $4.573~\mathrm{GHz}$\\
 Qubit frequency&   $\omega_q/2\pi$  & $\hbar \omega_q \sigz/2 $  & $6.181~\mathrm{GHz}$\\
 Tomography qubit frequency & $\omega_t/2\pi$ & $\hbar \omega_t \sigz^t/2$ & 3.459 GHz \\
 Readout frequency    & $\omega_r/2\pi$ & $\hbar \omega_r\hat{r}^\dagger \hat{r}$ & $7.875~\mathrm{GHz}$ \\
 Cavity-qubit cross Kerr rate    & $\chi/2\pi$ & $-\hbar \chi \adag\ahat \sigz$ & $5.25~\mathrm{MHz}$ \\
  Cavity-tomography qubit cross-Kerr rate& $\chi_{ct}/2\pi$  & $-\hbar \chi_{ct} \hat{a}^\dagger \hat{a} \sigz^t/2$   & 593 kHz \\
  Cavity self-Kerr rate& $\chi_{cc}/2\pi$  & $-\hbar \chi_{cc} \hat{a}^{\dagger 2} \hat{a}^2$   & 9 kHz \\
 \hline
  Circuit parameter& Symbol & Dissipation operator & Value \\
 \hline
 Qubit decay time & $T_{q}$  & $1/T_{q}\DD_{\sigm}$   & 22 ns \\
 Cavity decay time & $T_{c}$  & $1/T_{c}\DD_{\ahat}$   & 200 $\mu$s \\
 Tomography qubit decay time & $T_{t}$  & $1/T_{t}\DD_{\sigm^t}$   & 3.6 $\mu$s  \\
 Readout decay time & $T_{1, r}$  & $1/T_{1, r}\DD_{\hat{r}}$   &  415 ns \\
 Cavity dephasing time & $T_{c, \phi}$  & $2/T_{c,\phi}\DD_{\adag\ahat}$   & 36 $\mu$s \\
 Tomography qubit dephasing time & $T_{q, \phi}$  & $1/2T_{q, \phi}\DD_{\sigz^t}$   & 12.3 $\mu$s \\
\hline
\end{tabular}
\label{circuits parameters}
\end{table}

\subsection{Measurement setup}

The readout resonator, the tomography qubit, and the qubit are driven on resonance by pulses that are generated using an OPX from Quantum Machines®. It has a sampling rate of 1 GS/s. AWG driving pulses are modulated at a frequency 125 MHz for readout, 110 MHz for tomography qubit, 70 MHz for cavity and 68 MHz for the monitoring qubit. They are up-converted using I-Q mixers for the readout resonator, monitoring qubit and  cavity, regular mixers for the tomography qubit, with continuous microwave tones produced by three channels of an AnaPico® APUASYN20-4 for the readout resonator, cavity and qubit, Agilent® E8257D for tomography qubit. 
The fourth channel of AnaPico® APUASYN20-4 is used for the TWPA pump. An AnaPico® APSIN12G is used to provide a strong detuned drive on the qubit in order to calibrate the reflection coefficient~\cite{Stevens2022}.

The two reflected signals from the readout and qubit are combined with a diplexer and then amplified with a TWPA provided by Lincoln Labs~\cite{Macklin2015}. The follow-up amplification is performed by a HEMT amplifier from Low Noise Factory® at 4 K and by two room-temperature amplifiers. The two signals are down-converted using image reject mixers before digitization by the input ports of the OPX. We tune the TWPA pump frequency ($6.079~\mathrm{GHz}$) and power in order to reach a total quantum efficiency of $\eta=17 \pm 1$ \% at $\omega_q/2\pi + 10$~MHz. The efficiency was extracted from the measured mean and variance of the demodulated quadratures using the qubit as a reference for calibrating the gain of the amplification chain (see section 6 in Ref.~\cite{Essig2021}). However, we observed on posterior experiments that this efficiency can fluctuate by almost 20 \% between $\omega_q/2\pi$ and $\omega_q/2\pi - 9\chi$.

The frequency comb is generated by numerically computing the temporal shape corresponding to a frequency comb. The use of an I-Q mixer allows to use negative intermediate frequencies to expand the comb on both sides of the source frequency.

\begin{figure*}
\centering
\includegraphics[width=\linewidth]{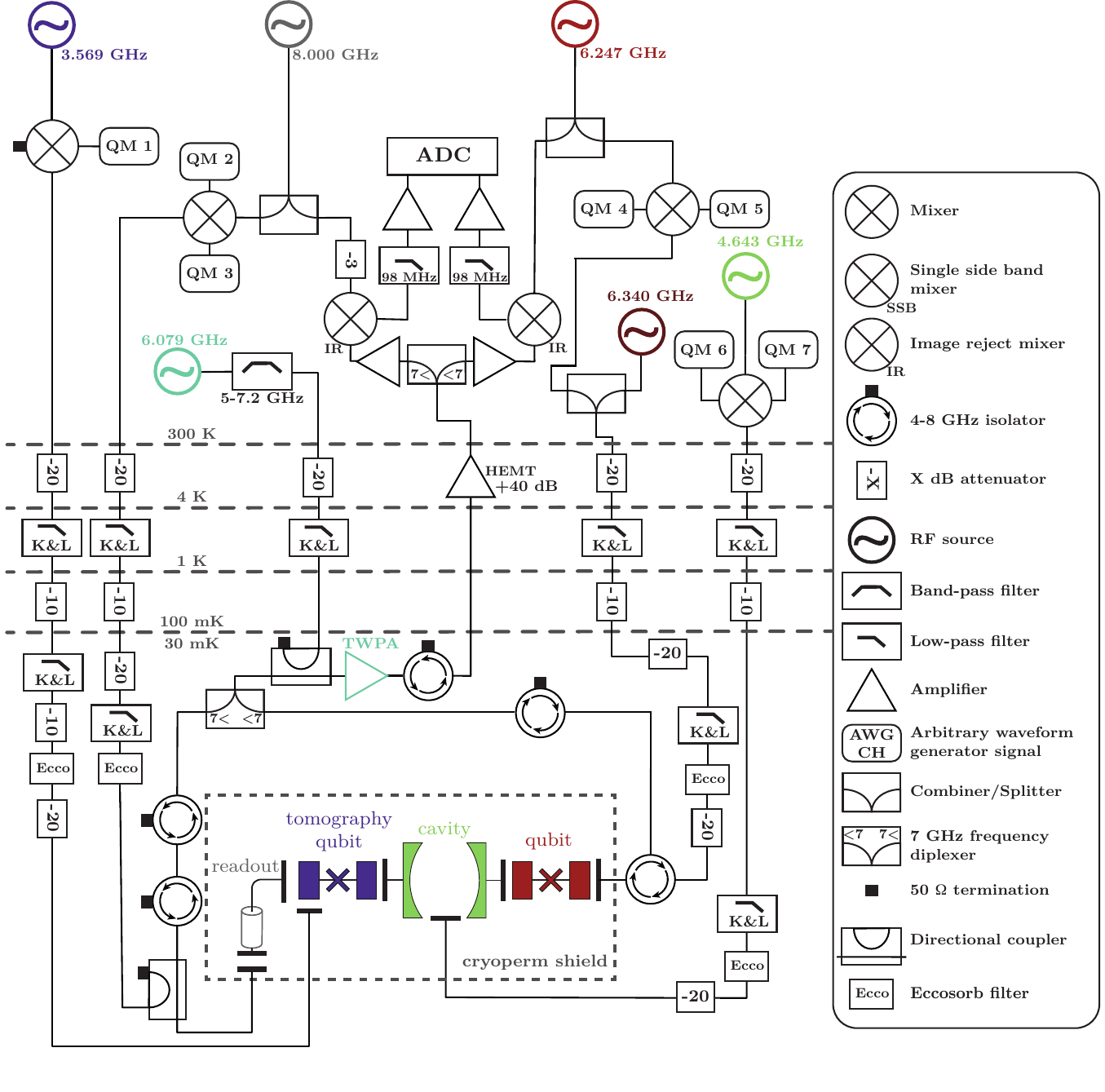}
\caption{Schematic of the setup. Each element has its RF source, whose color is matched. Room-temperature isolators are not represented for the sake of clarity.}
\label{fig:cablage}
\end{figure*}

\begin{figure}
\centering
\includegraphics[width=\linewidth*3/4]{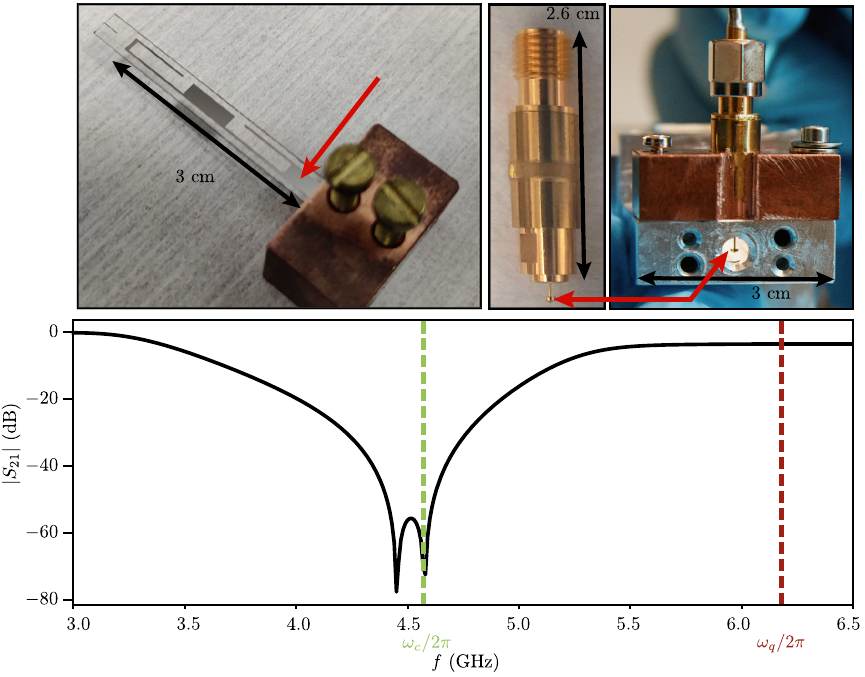}
\caption{Top panel: pictures of, from right to left, the qubit and the on-chip filter on its support, the pogo pin, and the pogo pin inserted in the tunnel in which the chip is inserted. Red arrows show the pogo pin and the contact point on the sample.
Bottom panel: frequency profile of the spurline filters, from an HFSS simulation. The actual frequency of the cavity is represented in green, that of the qubit in dark red. }
\label{fig:stub_pogo}
\end{figure}

\subsection{Comparison with previous experiment}
We summarize in table \ref{table_diff} the main differences with the reference \cite{Essig2021} that allowed us to reach single shot photon number measurement, extract a measurement rate and observe quantum jumps. One big difference is the use of a 3D cavity that extended the resonator decay by 2 orders of magnitude. Another key difference is the use of the analog of a Purcell filter for the cavity, galvanically connected in a 3D design. The filter allowed us to preserve the two orders of magnitude of improvement on the resonator lifetime despite a similar coupling rate between qubit and cavity. The expected measurement rate is slightly lower, because the dispersive coupling rate $\chi$ between the qubit and the cavity could have been larger to benefit from the strong qubit emission rate into the measurement line. But this is compensated by a far the longer decay time for the cavity. 

In terms of results, the main figure of merit of the experiment is $\Gamma_m T_c$, which was improved from 0.5 to 16, reaching the single shot regime. Additionally, the experimental technique changed, as we use a much wider comb experimentally, as it was proposed (but not implemented) in \cite{Essig2021}.

The expected measurement rate for the reference \cite{Essig2021} was not computed but is here obtained via the same model than for the blue curve of Fig.4a in the main text.

\begin{table}\caption{Summary of the differences with \cite{Essig2021}\label{table_diff}}
\begin{tabular}{ |p{6cm}||p{2cm}|p{2cm} | } 

  \hline
  & Work \cite{Essig2021}  & This work \\
 \hline 
 Architecture &  2D & 3D \\
 Cavity filter & no  & yes  \\
 Qubit decay rate & $42~\mathrm{ns}^{-1}$  &   $22~\mathrm{ns}^{-1}$   \\
 Dispersive shift $\chi/2\pi$ & $4.9~\mathrm{MHz}$  & $5.25~\mathrm{MHz}$   \\
 Cavity decay time $T_c$ &  3.8 $\mu$s & 240 $\mu$s \\
 Bandwidth of the comb &  $9 \chi$ &  $40 \chi$ \\
 Expected maximal measurement rate $\Gamma_m$ &  $130~\mathrm{ms}^{-1}$ &  $80~\mathrm{ms}^{-1}$ \\
 $\Gamma_m T_c$ &  $0.5$ &  $16$ \\
\hline
\end{tabular}
\end{table}

\section{Single-tone photon counting for pre- and post-selection}
\subsection{Single shot measurement with a single tone}
\begin{figure}
\centering
\includegraphics[width=\linewidth*4/5]{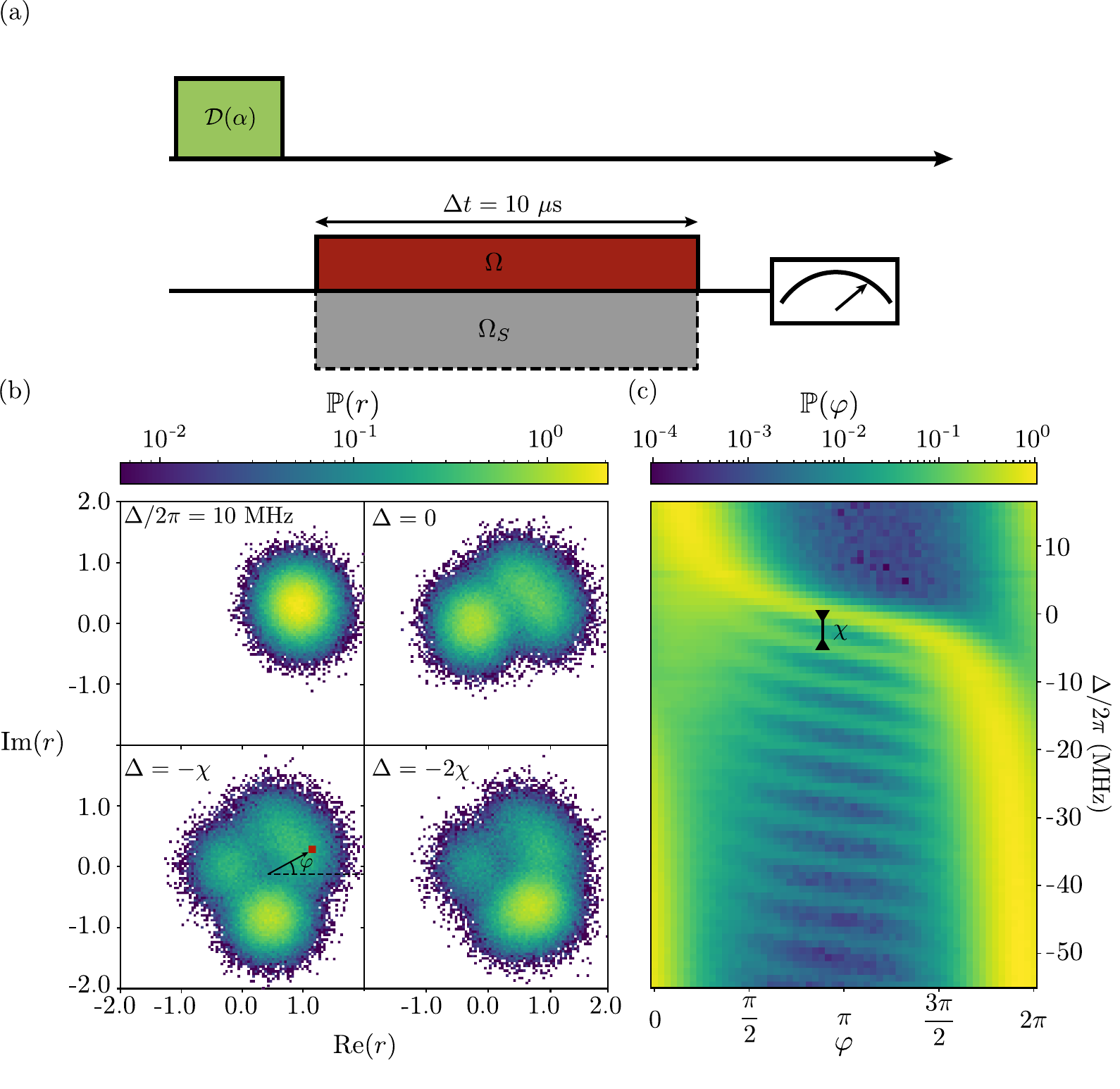}
\caption{(a) Pulse sequence. The cavity is driven on resonance so as to excite it with large number of photons, and a series of $10~\mathrm{\mu}$s readout pulse (red) with amplitude $\Omega$ and detuning $\Delta = \omega-\omega_q$ are applied on the qubit while the reflected signal is recorded to infer a reflection coefficient $r$. The same protocol is then performed while applying a strong ac Stark shift drive (gray, dashed). It detunes the qubit far from the frequencies of interest, which provides a reference signal to properly calibrate the reflection coefficient. (b) Histograms of the reflection coefficient $r$ constructed from this procedure. We can clearly distinguish the components coming from the different number of photons. (c)  Distribution of the argument $\varphi$ of the complex number $r$ computed from (b) as a function of the detuning $\Delta$. $\chi$ appears as the frequency translation parameter of the resonance of the qubit.}
\label{fig:blobs_chis}
\end{figure}

In this section, we focus on the case where the qubit is driven with a single tone to determine whether the cavity contains a certain number of photons~\cite{Gely2019,Essig2021}. This configuration is reciprocal to the usual dispersive readout of a qubit. 
We introduce the reflection coefficient as the ratio $\overline{r}=\langle\hat{a}_\mathrm{out}\rangle/\langle\hat{a}_\mathrm{in}\rangle$ in the steady state. The denominator is calibrated by applying a strong ac Stark shift drive to the qubit (Fig.~\ref{fig:blobs_chis}a), so that it is detuned far from $\omega_q$. We generalize the definition of a reflection coefficient $r$ to the case of single measurement records. Therefore, after a displacement of the cavity, we can build histograms of $r$ as a function of the frequency of the tone (Fig.~\ref{fig:blobs_chis}b). For a given photon number, the average reflection coefficient spans a circle in the complex plane as the probe frequency is changed~\cite{thesisessig2021}. We introduce a phase $\varphi$ relative to the center $r_0=0.3+0\times i$ of that circle (see Fig.~\ref{fig:blobs_chis}b). We recover the usual $2\pi$ phase shift of the reflection coefficient on a harmonic oscillator in the angular distribution $\mathbb{P}(\varphi)$. Each number of photons leads to a unique phase shift as can be observed in Fig.~\ref{fig:blobs_chis}c. In particular, when we drive at $\omega_q-n \chi$, the histogram peak corresponding to Fock state $|n\rangle$ is centered on a $\pi$ phase shift. Thus, even though the histograms of various photon numbers overlap, it is possible to pre/post-select this Fock state by setting a threshold on the real value of one averaged single record. The tighter the threshold, the higher the fidelity but the smaller the selection yield as we discuss now.

\subsection{Pre- and post selection using the qubit}
\begin{figure*}  
\centering
\includegraphics[width=\linewidth*2/3]{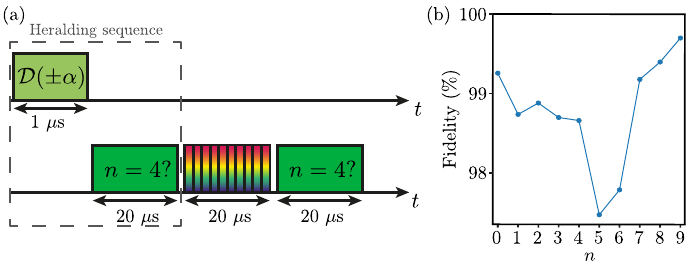}
\caption{(a) Pre- and post-selection pulse sequence. The cavity is alternatively displaced by $\pm \alpha = 3.4$ (which is believed to improve slightly the pre-selection yield without reset compared to $+\alpha$ always). A first pre-selection $20$ µs pulse is then applied to the qubit. This heralding sequence (dashed box) is repeated until the measured $\mathrm{Re}(r)$ gets below a threshold. A 20~µs comb pulse is then applied to the qubit. In practice, it results from the concatenation of 10 identical sequences of 2~µs. We then apply a post-selection pulse on the qubit. (b) Estimated pre-selection fidelity as a function of the Fock state. The selection yields associated to the threshold we used (red dashed lines in Fig.~\ref{fig:fock_state_selection}) are given in Table~\ref{table_yield}. These fidelities are determined by computing the probability to get the right number of photons after applying the threshold. For this, a prior probability distribution for the number of photons is needed, which we deduced from the yield statistics in Table~\ref{table_yield}.}
\label{fig:selection_sequence}
\end{figure*}

\begin{figure*}
\centering
\includegraphics[width=\linewidth]{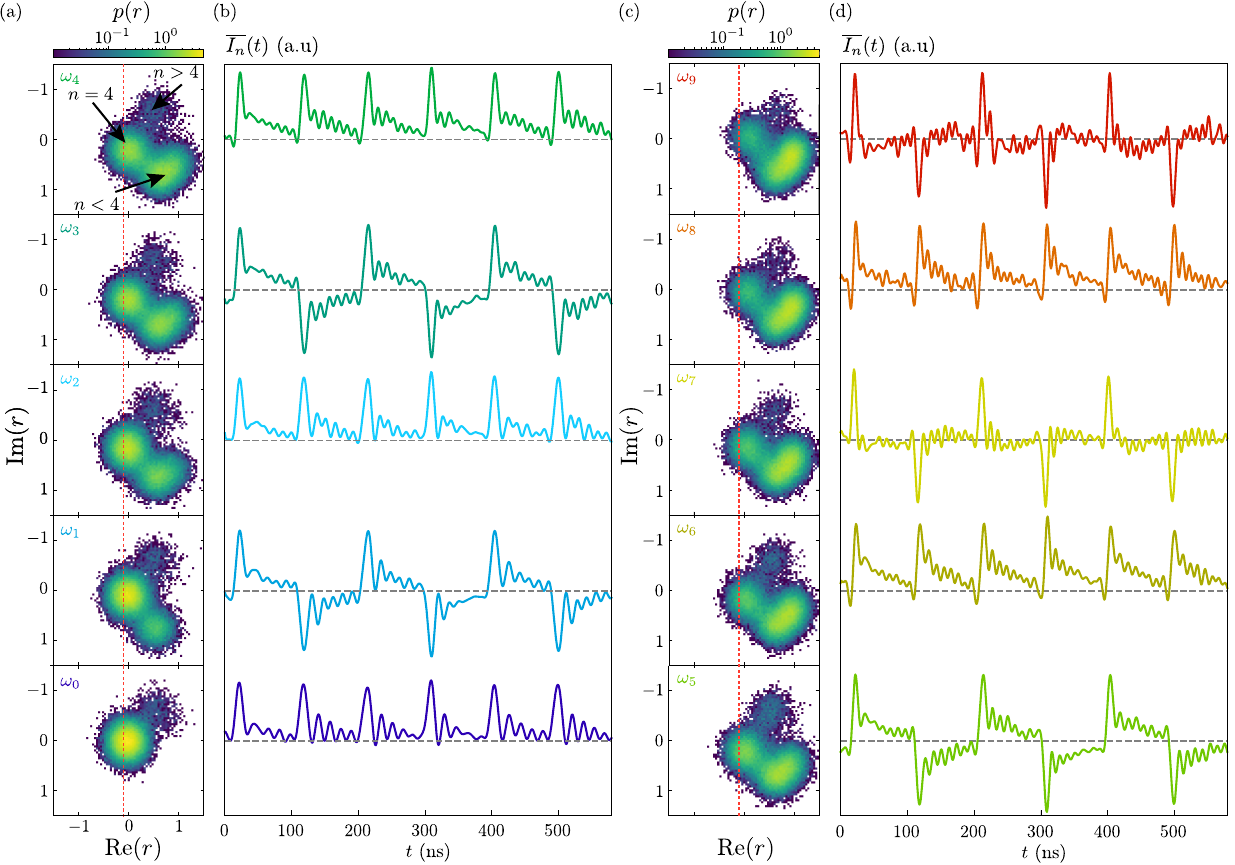}
\caption{(a) and (c) Histograms of the measured quadratures during the second (post-selection) pulse on the qubit for each number $n$ of photons corresponding to a driving tone and heterodyne measurement at $\omega_n$. The upper threshold on $\mathrm{Re}(r)$ for the pre- and post-selection readout is materialized by the dashed red lines. There are more than one Gaussian component due to the decay between the pre- and post-selection pulses.   (b) and (d) Reconstructed average fluorescence of the qubit for each number of photons. Note that the figure (b) is identical to Fig.~2e.}
\label{fig:fock_state_selection}
\end{figure*}

The protocol we use to pre- and post-select a given number of photons is shown in Fig.~\ref{fig:selection_sequence}a. It is performed by sending a $20~\mu$-drive at $\omega_q-n\chi$ before and after the measurement. For each pulse, we determine the associated reflection coefficient. The measurement is heralded to Fock state $|n\rangle$ if both measurements give reflection coefficients falling below threshold. The thresholds we use are shown in Fig.~\ref{fig:fock_state_selection}a,c as dashed red lines on the measured histograms at the post-selection step for each desired photon number $n$. The value of the threshold $r_0$ was chosen so that the likelihood ratio between \{the number of photons $n$ being equal to the selected number of photons $k$\} and \{$n$ being equal to $k'$\} is higher than 100 for every target photon number $k$(i.e., for any selected frequency $f-k\chi$) and for any $k'\neq k$:
$$\frac{\mathds{P}(\Re(r)<r_0|n = k)}{\mathds{P}(\Re(r)<r_0|n = k’)}>100.$$

There is a trade-off between preparation fidelity by heralding and its yield. We choose a drastic pre-selection threshold on $\mathrm{Re}(r)$ that ensures preparation fidelities of at least 97~\% by the pre-selection pulse alone as shown in Fig.~\ref{fig:selection_sequence}b while the selection yields (below 1~\%) are reported in Table~\ref{table_yield}.

\begin{table}\caption{Yield table of the pre- and post-selection after a displacement of $\alpha\approx 3.4$ when the kick angle $\theta$ is set to 0.\label{table_yield}}
\begin{tabular}{ |p{2cm}||p{2cm}|p{2cm} |p{5cm}| } 

  \hline
 Fock state & Pre-selection  & Post-selection & Typical number of samples $2N_n$ \\
   & fraction (\%) & fraction (\%) & after post-selection\\
 \hline 
 0 &  0.5  & 33.3 & 200580 \\
 1 & 0.7  & 26.0  & 157370\\
 2 &  0.7 & 19.7 & 119370\\
 3 & 0.6 & 14.1 & 85800\\
 4 & 0.5 & 11.1 &67150\\
 5 & 0.4  & 7.4  & 44990\\
 6 & 0.3  & 4.2  &25650\\
 7 & 0.2  & 3.2  &19330\\
 8 & 0.2  & 1.9  &11360\\
 9 & 0.1  & 0.9  & 5580\\

\hline
\end{tabular}
\label{yield table}
\end{table}

\section{Computing the theoretical measurement rates}
\label{sec:theoretical_rates}

Our model for the measurement focuses on what happens after a single kick of the qubit state. Indeed, the experiment can be decomposed into a repetition of the same sequence. Every $\pi/\chi$, the qubit is prepared in a certain state $\rhohat_q$ that depends on the amplitude of the comb, and decays. Since in our case, $\pi/\chi\simeq 4T_{q}$, we use a model where the qubit fully relaxes to its ground state between each kick. We can thus assume the kick places the qubit in a pure state $\rhohat_q = \ketbra{\psi}$ with $\ket{\psi} = \cos(\theta/2)\ket{0} +\sin(\theta/2)\ket{1}$, where $\theta = 2\pi \Omega/2\chi$ is the kick angle (see main text).

Following section F of the supplemental material of Ref.~\cite{Clerk2010IntroductionAmplification}, the measurement rate of the photocounting can be extracted from the mutual information between the qubit frequency and a measurement record, for the most entropic prior (all states assumed equally likely initially). We compute it in the case where the qubit can only have one out of two frequencies separated by $\chi$ corresponding to a given photon number or one more.

\begin{figure}
\centering
\includegraphics[width=\linewidth]{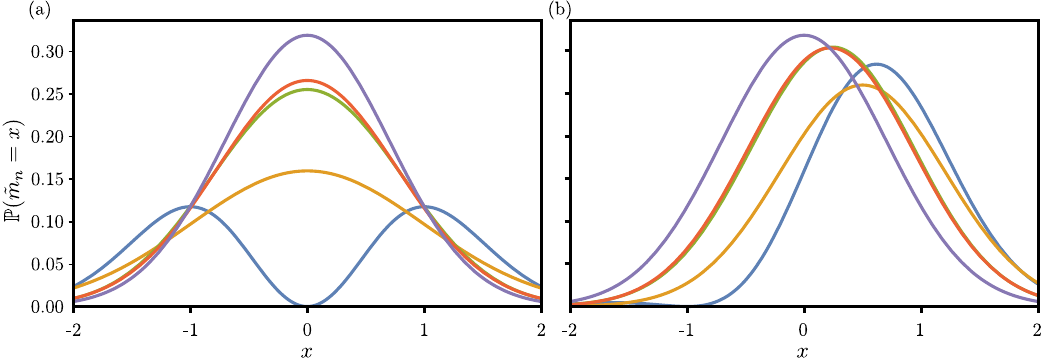}
\caption{Calculated probability distributions of $\mathbb{P}(\tilde{m}_n=x)$ as a function of $x$ when the cavity is in $|n\rangle$ for (a) $\ket{\psi} = \ket{1}$ and (b) $\ket{\psi} = (\ket{0}+\ket{1})/\sqrt{2}$ . Blue: $\eta = 1$. Orange: its Gaussian approximation with same mean and covariance matrix (on the whole complex plane $\tilde{m}_n\in\mathbb{C}$).  Green: $\eta = 0.2$. Red: its Gaussian approximation. Purple: case where $\ket{\psi} = \ket{0}$.}
\label{fig:gaussian_approx}
\end{figure}

We model the experiment as follows. The bipartite system  \{cavity, qubit\} starts in the initial state $\hat{\rho}(0)=\brho_c\otimes \brho_q$.
The full Stochastic Master Equation (SME), neglecting the pure dephasing of the qubit, of the system reads
\begin{multline}
  d\brho = - i \left[ \wc \adag\ahat + \tfrac{1}{2}\omega_q \Sz - \tfrac{1}{2} \chi \adag\ahat \Sz , ~\brho    \right] \mathrm{d}t
  + \left( \frac{1}{T_q}  \DD_{\Sm}(\brho) + \frac{1}{T_c} \DD_{\ba}(\brho) \right) \mathrm{d}t
  \\ + \sqrt{\frac{\eta}{T_q}} \Big(\Sm \brho - \Tr{\Sm \brho} \brho  \Big) \mathrm{d}W^* + \sqrt{\frac{\eta}{T_q}} \Big(\brho \Sp - \Tr{\brho \Sp} \brho  \Big)\mathrm{d}W
\end{multline}

where $\mathrm{d}W$ is now a complex Wiener process such that $\mathrm{d}W\mathrm{d}W^* = \mathrm{d}t$. The complex measurement record with efficiency $0<\eta<1$ reads $$ \mathrm{d}y_t = \sqrt{\eta/T_q} \Tr{\brho\Sm}  \mathrm{d}t  + \mathrm{d}W. $$

It can be proven that the information contained in the measurement record $y_t$ about the photon number in the cavity is preserved by demodulating the measurement record as follows
\begin{align}
\tilde{m}_k = \frac{1}{\sqrt{T_q}}\int_0^{+\infty}e^{i (\omega_q - k\chi)t-\frac{t}{2 T_q}}\mathrm{d}y_t,\label{eq:S2}
\end{align}
for all possible photon numbers $k$.

We can take into account the finite efficiency $\eta$ of the detection as a beam-splitter that combines a fraction of the field emitted by the qubit with vacuum noise. Assuming the cavity is in state $|n\rangle$, the outcome $\tilde{m}_n$ is a stochastic variable that reads
\begin{align}
\tilde{m}_n = \sqrt{\eta}\alpha+ \sqrt{1-\eta} \gamma,
\end{align}
where $\alpha$ and $\gamma$ are stochastic variables whose probability densities are respectively given by the Husimi Q functions of the qubit state $\rhohat_q$ and of the vacuum. 
Owing to the low measurement efficiency $\eta\approx 0.2$, the distribution function of $\tilde{m}_n$ is almost Gaussian as can be seen in Fig.~\ref{fig:gaussian_approx} in the cut along the real axis. For the quantum efficiency $\eta \leq 0.2$ of the experiment, we see that the two curves are very close, thus justifying the Gaussian approximation of the measurement outcomes.

The other measurement outcomes $\tilde{m}_{k\neq n}$ are obtained with a wrong demodulation frequency $\omega_q-k\chi$. However there is still a finite overlap $\frac{\Gamma_q}{\Gamma_q + i (k-n)\chi}$ between the decaying modulation functions at $n$ and $k$, where $\Gamma_q=1/T_q$. The outcome $\tilde{m}_{k\neq n}$ can thus be written as a combination of $\tilde{m}_n$ and of a stochastic variable $\beta_k$ distributed as the Husimi Q function of the vacuum state with weights corresponding to that overlap.
\begin{align}
&\tilde{m}_{k\neq n}= \frac{\Gamma_q}{\Gamma_q + i (k-n)\chi}\tilde{m}_n + \frac{i(k-n)\chi}{\Gamma_q + i (k-n)\chi}\beta_k
\end{align}

Knowing that $\alpha, \beta_k, \gamma$ are independent variables, one can compute $\mathbb{P}(\vec{\tilde{m}}|n)$. Knowing that the repetition time is $\tau = \frac{\pi}{\chi}$, we use equation (4) in the main text to predict the measurement rates for $\eta = 0.17, 0.2$ (blue in Fig.~4a) and $\eta = 1$ (blue in Fig.~4b). For larger kick angles, we observe a discrepancy between the experimental data and the model in Fig.~4a. The x-axis  corresponds to the ideal angle $\theta$ reached by the qubit for an infinite frequency comb. However, both in the experiment and in the simulation, the finite duration of the kicks lowers the actually reached rotation angle below $\theta$. This is why the blue curve behaves as expected while the experiment and simulations appear distorted in $\theta$.

\section{Data processing}
This section provides details on all the steps involved in the data processing used in the letter.

\subsection{Quantum model of the recorded voltage}

The qubit couples the incoming modes $\hat{a}_{\mathrm{in}}(t)$ to the outgoing modes $\hat{a}_{\mathrm{out}}(t)$ following the input-output relation
\begin{equation}
    \hat{a}_{\mathrm{out}}(t)= \hat{a}_{\mathrm{in}}(t) - \sqrt{\frac{1}{T_q}}\hat{\sigma}_-(t).
\end{equation}

\begin{figure*}[b]
\centering
\includegraphics[width=\linewidth*2/3]{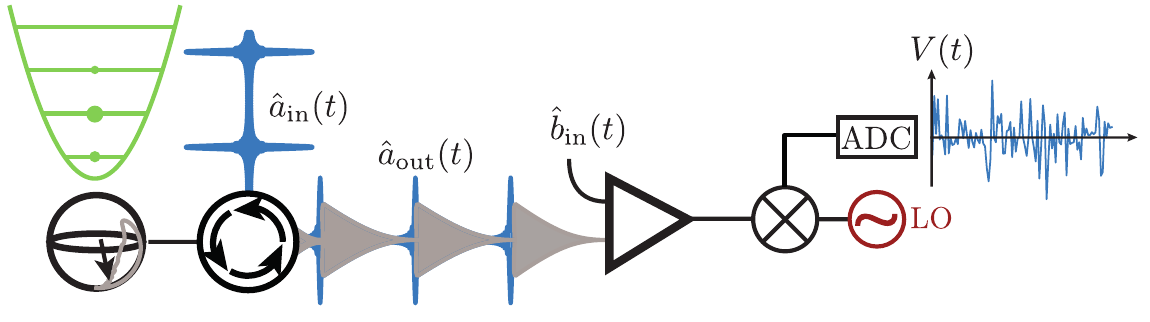}
\caption{Scheme of the quantum model of the detection of the fluorescence.}
\label{fig:scheme}
\end{figure*}

The outgoing modes $\hat{a}_{\mathrm{out}}(t)$ get amplified and downconverted by a local oscillator at a frequency $\omega_{LO}/2\pi = 6.247~\mathrm{GHz}$. The noise added  by the amplifiers can be modeled as a finite temperature of input modes $\hat{b}_\mathrm{in}(t)$, which are assumed to be in a Gibbs state. This model (Fig.~\ref{fig:scheme}) leads to the following expression for the voltage $V(t)$ that is measured by the analog-to-digital converter (ADC) at room temperature, where $G$ encompasses the conversion factor between the $\sqrt{\mathrm{Hz}}$ of the propagating mode amplitudes and Volts of the ADC records. $V(t)$ is the measurement outcome of the observable 
\begin{equation}
    \hat{V}(t) = \sqrt{G}\Big(\hat{a}_\mathrm{out}(t)e^{i\omega_\mathrm{LO}t} + \hat{a}^\dagger_\mathrm{out}(t)e^{-i\omega_\mathrm{LO}t}\Big)
+ \sqrt{G-1}\Big(\hat{b}^\dagger_\mathrm{in}(t)e^{i\omega_\mathrm{LO}t} + \hat{b}_\mathrm{in}(t)e^{-i\omega_\mathrm{LO}t}\Big).\label{eq:hatVoft}
\end{equation}

\begin{figure*}
\centering
\includegraphics[width=\linewidth*3/5]{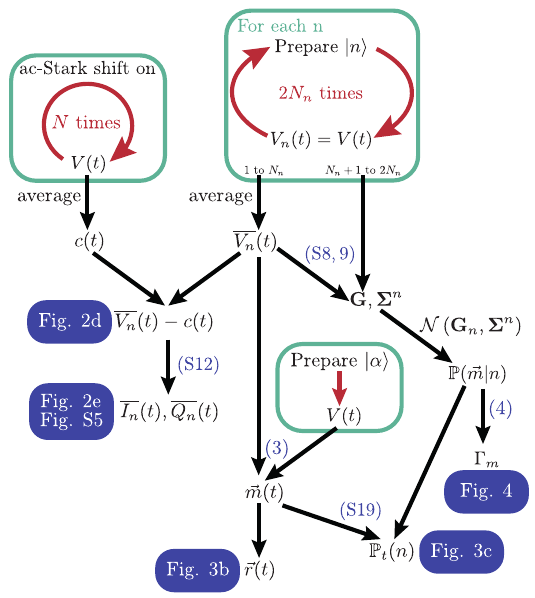}
\caption{Diagram of the signal processing (see text).}
\label{fig:diagram}
\end{figure*}

\subsection{Data processing flow}

From the measured voltages, we get several relevant quantities that are discussed in the main text. Here, we show how they are related. Each Fock state $\ket{n}$ from $0$ to $9$ is prepared a large number of times (see Table. \ref{yield table}) by pre- and post-selection, leading to as many measurement records $V_n(t)$ of the voltage $\hat{V}(t)$ over 10 slices of $2~\mathrm{\mu s}$.

For any measurement record $V(t)$, we define the measurement outcome $\vec{m}$ whose components are\begin{equation}
m_j = \int_{0}^{2~\mu\mathrm{s}} V(t) \overline{V_j}(t)\mathrm{d}t,
\end{equation}
which corresponds to $m_j(2~\mu\mathrm{s})$ in Eq.~(3) of the main text.

For each photon number $n$, we compute the mean $\overline{V_n}(t)$ of the $V_n(t)$ whose expectation is $\langle \hat{V}(t)
\rangle_{|n\rangle}$. We can then calculate the Gram matrix (matrix of scalar products)
\begin{equation}
\mathbf{G}_{nm} = (\overline{V_n}|\overline{V_m}) = \int_0^{2~\mu s}\overline{V_n}(t) \overline{V_m}(t)dt.
\end{equation}
Since the number of samples is finite, the mean $\overline{V_n}$ is a stochastic variable. In order to avoid biasing the Gram matrix with the particular realization of $\overline{V_n}$, we compute each scalar product between averages $(\overline{V_n}|\overline{V_m})$ by taking half the samples for the left part and the other half for the right part (see Fig.~\ref{fig:diagram}). We also compute the covariance matrices $\mathbf{\Sigma}^n$ of the probability distribution of the measurement outcomes $\vec{m}^n$ obtained from $V_n(t)$ when there are $n$ photons. Its matrix elements read

\begin{equation}
\mathbf{\Sigma}^n_{jk} = \mathrm{Cov}\{(V_n|\overline{V}_j), (V_n|\overline{V_k})\} = \overline{(m_j^n-\mathbf{G}_{jn})(m_k^n-\mathbf{G}_{kn})}.
\end{equation}
As for the Gram matrix, we compute each scalar product $(V_n|\overline{V}_j)$ by taking half the samples for the left part and the other half for the right part.

As explained in Fig.~\ref{fig:gaussian_approx}, the distribution probability $\mathbb{P}(\vec{m}|n)$ of $\vec{m}^n$ can be approximated by a Gaussian law for our low efficiency $\eta\simeq 0.2$. Therefore, we use the experimentally determined Gram and covariance matrices to reconstruct this law $\mathcal{N}(\mathbf{G}_n, \mathbf{\Sigma}^n)$. It is used in the Bayesian update of the photon number and for the computation of the measurement rate. For the estimation of $\mathds{P}(\vec{m}|n)$ used in the Bayesian filter for the photon number tracking in Fig.~3, we pre- and post-selected $60600\times10$ realizations following the protocol in Fig.~\ref{fig:selection_sequence}. However, since we require many drive amplitudes for Fig.~4, we selected only $6600\times10$ realizations for each amplitude in that figure. We there chose to remove the numbers $8$ and $9$ from the analysis as the insufficient number of samples induces very large uncertainties on the measurement rate.

\begin{figure*}
\centering
\includegraphics[width=\linewidth*1/2]{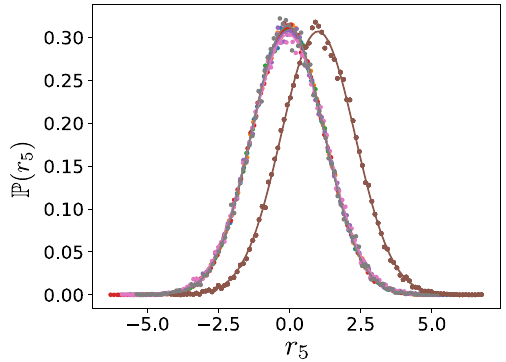}
\caption{Brown dots: measured probability distribution for the record $r_5$ after 2 $\mu$s of integration when 5 photons are prepared in the cavity. Dots with other colors: measured probability distribution for the record $r_5$ when there are $k$ photons in the cavity, for $k$ between 0 and 9, $k\neq5$, and the same integration time. Solid line: gaussian fits.}
\label{fig:basis5}
\end{figure*}

From the $m_n$, we can compute the $r_n$ (as in the main text) and look at their probability distribution, which is shown Fig. \ref{fig:basis5} for $r_5$. The average value of $r_5$ is 1 when there are $5$ photons in the cavity (brown), and 0 otherwise. All the records that are conditioned to other numbers of photons have the same Gaussian law, centered around 0. Because of the noise of the measurement, the value of the records may take values as large as 8 or as low as -8, which can also be seen in Fig.~3b. This is analogous to the measurement record of a dispersive readout of a qubit. Note here that all the $r_n$ are correlated in general, so looking at the marginals is not enough to extract all the information about the number of photons. In particular, while this figure gives a hint about how well we can distinguish the Fock state 5 from the others in $2 ~\mu s$, it only gives a lower bound on the Signal-to Noise Ratio.

\subsection{Reconstruction of the qubit fluorescence signal}

In order to avoid being blinded by the drive, we remove the comb contribution (which is measured by driving the qubit off resonance using  ac-Stark effect)
\begin{equation}
c(t) = \sqrt{G}\Big(\langle\hat{a}_\mathrm{in}(t)\rangle e^{i\omega_\mathrm{LO}t} + \langle \hat{a}^\dagger_\mathrm{in}(t)\rangle e^{-i\omega_\mathrm{LO}t}\Big)\label{eq:coft}
\end{equation} to the average measurement records $\overline{V_n}(t)$ (Fig.~2d in the main text).

Following Eqs.~(\ref{eq:hatVoft}) and (\ref{eq:coft}), we express the expectation value of $\overline{V_n}(t)-c(t)$ as
\begin{equation}
\overline{V_n}(t)-c(t) = -\sqrt{\frac{G}{T_q}} \Big(\langle\hat{\sigma}_-(t)\rangle_{\ket{n}} e^{i\omega_\mathrm{LO}t}+ \langle\hat{\sigma}_+(t)\rangle_{\ket{n}}e^{-i\omega_\mathrm{LO}t}\Big).
\end{equation} 
Here, $ \langle\hat{\sigma}_-(t)\rangle_{\ket{n}}$ oscillates at $(\omega_q-n\chi)$, so $\langle\hat{\sigma}_-(t)\rangle_{\ket{n}}e^{i\omega_\mathrm{LO}t}$ oscillates at $\omega_\mathrm{LO}-\omega_q+ n\chi = \omega_\mathrm{IF}+  n\chi$. To reconstruct the fluorescence of the qubit in the frame rotating at its frequency, we bring these oscillating signals to zero frequency.
This is performed using their Hilbert transform $\mathcal{H}$ to reconstruct their analytic form $-\sqrt{\frac{G}{T_q}} \langle\hat{\sigma}_-(t)\rangle_{\ket{n}} e^{i\omega_\mathrm{LO}t}$ (here with negative frequencies). This allows to perform a sideband-free numerical demodulation to recover the average coherence of the qubit. It yields 
\begin{equation}\label{hilbert transform}
    \overline{I_n}(t) - i \overline{Q_n}(t) = e^{-i(\omega_\mathrm{IF}+n \chi)t + i\phi_n}\Big(\overline{V_n}(t)-c(t) + i \mathcal{H}[\overline{V_n}-c](t)\Big)
\end{equation}
where $\phi_n$ is chosen such that the most significant quadrature, i.e the one carrying the most fluorescence, is $\overline{I_n}$. These quadratures are plotted in Fig.~2.e in the main text and Fig.~\ref{fig:fock_state_selection}. Note that the demodulation frequency is $(\omega_\mathrm{IF}+n\chi)/2\pi$ and not $-n\chi$, which originates from the fact that the qubit is driven with a lower sideband.

\subsection{Relevance of the demodulation procedure}

In this section, we justify how well the demodulation we perform following Eq.~(3) (main text) preserves the information about the photon number that is contained in the measurement records. Indeed, using the average measurement record $\overline{V_n}(t)$ as a demodulation function does not straightforwardly appear equivalent to  Eq.~(\ref{eq:S2}) that states that using the demodulation function $e^{i (\omega_q - n\chi)t-\frac{t}{2 T_q}}$ preserves all the information about the photon number. 

Actually, the demodulation function $e^{i (\omega_q - n\chi)t-\frac{t}{2 T_q}}$ is proportional to $\Tr(\overline{\brho_n(t)}\Sm)$, where $\overline{\brho_n(t)}$ is the solution of a simple Lindblad master equation describing the evolution of a qubit at frequency $\omega_q - n\chi$ and with a relaxation time $T_q$. Therefore a valid way to demodulate the measurement record in the experiment would be to measure the average down-converted and pre/post-selected complex amplitude $e^{i\omega_{\mathrm{LO}}t}\Tr(\overline{\brho_n(t)}\Sm)$ first, and to use it to demodulate the measurement record. Formally, it corresponds to $\overline{V_n}(t)-c(t)+ i\mathcal{H}[\overline{V_n}-c](t)$ defined in Eq.~(\ref{hilbert transform}), where $\mathcal{H}$ is the Hilbert transform. 
However, using $\overline{V_n}(t)$ alone does not lead to any visible loss of information between the two methods. As a matter of fact, demodulating by $\overline{V_n}(t)-c(t)+ i\mathcal{H}[\overline{V_n}-c](t)$, $\overline{V_n}(t)-c(t)$ alone or $\overline{V_n}(t)$ does gather the same amount of information, up to experimental uncertainties. 

Note that this simplification is only possible thanks to the relatively low quantum efficiency $\eta$.
 Were the efficiency $\eta$ close to $1$, we would be able to gather information in all cumulants of the probability distribution of $\tilde{m}$ as well (see Fig.~\ref{fig:gaussian_approx}), which would require to keep the full expression $\overline{V_n}(t)-c(t)+ i\mathcal{H}[\overline{V_n}-c](t)$. 

\subsection{Measurement rate estimation}
Once $\mathbf{G}$ and $\mathbf{\Sigma}$ known, the measurement rate can be computed using
\begin{equation}\label{MI_hetero}
\begin{split}
\tau\Gamma_m  = I(n:\vec{m}) =&-\int \mathbb{P}(\vec{m})\log\mathbb{P}(\vec{m})\mathrm{d}\vec{m}\\
&+\sum_n \mathbb{P}_0(n)\int \mathbb{P}(\vec{m}|n)\log\mathbb{P}(\vec{m}|n)\mathrm{d}\vec{m},
\end{split}
\end{equation}
for any prior $\mathbb{P}_0$ for which there exists a number $q$ such that $\mathbb{P}_0(q)=\mathbb{P}_0(q+1)=1/2$, and where $I(n:\vec{m})$ is the mutual information of $n$ and $\vec{m}$. 

A subtlety arises from the fact that this expression is only valid for times $\tau$ such that $\tau\Gamma_m\ll 1$. For the largest measurement rates we obtained in the experiment, we cannot use the full integration time of $2~\mu\mathrm{s}$. However it is straightforward to compute what this expression gives if $\tau$ were as small as the period of the pulse train driving the qubit, which is here $\pi/\chi=\tau/21$. For this computation, we then rescale the Gram matrix into $\mathbf{G}' = \mathbf{G}/\sqrt{21}$.

Note that we chose to integrate over $2~\mathrm{\mu s}$ because of a technical reason. It is simply the smallest multiple of the pulse train period that is also a multiple of the inverse sampling rate of the OPX instrument (1 ns).

The mutual information obtained in this way is actually biased. Indeed, even if all the $\mathbf{G}_n$ and $\mathbf{\Sigma}^n$ are equal ($I(n:\vec{m}) =0$), their experimental value are slightly different owing to noise, thus rendering $I(n:\vec{m})$ positive. To estimate this bias, we measured $I(n:\vec{m}_0)$ for zero driving amplitude, and assumed that it should be zero exactly, neglecting the thermal emission of the qubit. We observed that $I(n:\vec{m}_0)$ scales linearly with the inverse of the number of samples. We then used the computed $I(n:\vec{m}_0)$ at zero amplitude as a calibration of the systematic errors for all the other amplitudes. For each driving amplitude, the expected positive systematic error inversely scales with the number of measured samples given in Table \ref{yield table}. The error bars displayed in Fig.~4 correspond to this systematic error.

\subsection{Simulation}
In order to simulate the experiment, we numerically solve the following Stochastic Master Equations (SME), for each $n$ between $0$ and $7$.

\begin{equation}\label{SME}
    \begin{split}
        \mathrm{d}\rhohat(t) = -\frac{i}{\hbar}[\hat{H}_n, \rhohat]\mathrm{d}t + \frac{1}{T_{q}}\DD_{\sigm} (\rhohat(t))\mathrm{d}t + \sqrt{\frac{\eta}{T_{q}}}\cH_{\sigm}(\rhohat(t))\mathrm{d}W(t),
    \end{split}
\end{equation}
with
\begin{align}
    \hat{H}_n &= \hbar \left(\omega_{IF} + n \chi\right) \ketbra{e}\\
    \DD_{\hat{A}} (\rhohat) &= \hat{A}\rhohat\hat{A}^\dagger
    -\frac{1}{2}\Big(\hat{A}^\dagger\hat{A}\rhohat+\rhohat\hat{A}^\dagger\hat{A}\Big)\\
    \cH_{\hat{A}}(\rhohat) &= \rhohat\hat{A}^\dagger + \hat{A}\rhohat - \Tr{\rhohat\hat{A}^\dagger + \hat{A}\rhohat}.
\end{align}

The homodyne measurement record reads
\begin{equation}
        \mathrm{d}y_t = \sqrt{\frac{\eta}{T_{q}}}\Tr(\sigx\rhohat(t))\mathrm{d}t + \mathrm{d}W(t),
\end{equation}
where $\mathrm{d}W$ is a Wiener process with zero mean and variance $\mathrm{d}W^2=\mathrm{d}t$. The record models the outcome of a \textit{homodyne} detection of the fluorescence of the qubit for a known number of photons $n$ in the cavity, with a quantum efficiency of $\eta$. The intermediate frequency is $\omega_{IF}/2\pi = 66$~MHz.

The orange curve of Fig.~4a results from calculating 10 000 quantum trajectories solutions of these SME for each amplitude and frequency, for $\eta = 0.17$ and $0.2$ and applied the same procedure as for the experimental records for $V(t)=\sqrt{G}\mathrm{d}y_t/\mathrm{d}t$. The only difference is that the average signal $\overline{V}$ was directly obtained as the solution $\rhohat(t)$ to the Lindblad equations, given by Eq.~(\ref{SME}) when $\eta = 0$, and computing $\sqrt{\frac{\eta G}{T_{q}}}\Tr(\sigx\rhohat(t))$.

\subsection{Bayesian filter for the photon number}

Tracking the photon number corresponds to actuating the probability of having $n$ photons in the cavity $\mathds{P}_t(n)$ conditioned on the measurement record $V$ from time 0 to $t$. Given $\mathds{P}_{j\tau}(n)$ and the measured $\vec{m}(j\tau+\tau)$, we compute $\mathds{P}_{j\tau+\tau}(n)$ as follows: 
\begin{itemize}
    \item During $\Delta t =2$ $\mu$s, apply the dissipation operator $1/T_c \DD_{\hat{a}}$ on the diagonal density matrix of the cavity, whose diagonal elements are the $\mathds{P}_{j\tau}(n)$ coefficients. We denote the resulting diagonal elements of the density matrix as $\mathds{P}'_{j \tau}(n)$, and use a cavity lifetime $T_c = 200~\mu$s,
    \item Actuate it with the corresponding $\vec{m}(j\tau+\tau)$ following
    \begin{equation}
    \mathds{P}_{j\tau+\tau}(n)=\mathds{P}'_{j \tau}(n) \mathds{P}(\vec{m}(t)|n)/Z,\label{eq:Bayesian_update}
    \end{equation}
    where $Z$ is a normalization constant ensuring $\sum_n \mathds{P}_{j\tau+\tau}(n) = 1$.
\end{itemize}

\subsection{Average time to reach a given confidence on the photon number measurement}

\begin{figure*}[h!] 
\centering
\includegraphics[width=\linewidth*1/2]{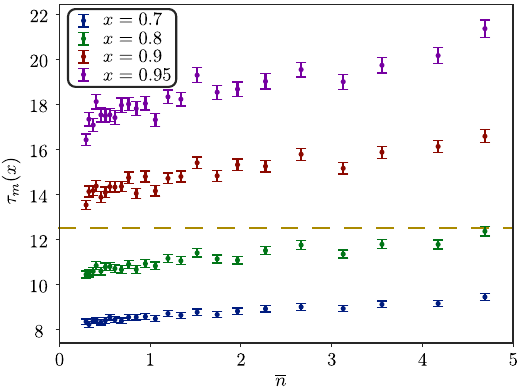}
\caption{Dots: average measurement time $\tau_m(x)$ required so that one of the photon number probabilities $P_n(t)$ exceeds the confidence $x=70~\%$ (blue), 80~\% (green), 90~\% (red) or 95~\% (purple). The times are represented as a function of the mean photon number at time $0$. Error bars represent the uncertainty coming from the finite number 1100 of realizations that were processed. Dashed line: Measured $1/\Gamma_m$ under the same conditions of driving ($\theta=\pi/2$ as in Fig.~3).}
\label{fig:tchar}
\end{figure*}
In addition to the measurement rate, it is possible to extract the average time it takes for the measurement procedure to reach a certain confidence $x<1$ in the number of photons (Fig.~\ref{fig:tchar}). Using the same set of measurement records from which the realization in Fig.~3 is taken, and starting at an arbitrary time $t_0$ with an agnostic probability vector $\mathds{P}_{t_0}(n)= 1/10$, we  process the data with the Bayesian update \textit{without including the dissipation}, until the time $t_0+\tau$ at which the estimator reaches a confidence $x$ for one of the numbers of photons. We then define $\tau_m(x) = \overline{\tau}(x)$, the average time that is required to reach the level of confidence $x$.
Using the average expected number of photons $\overline{n}$ (see Fig. \ref{fig:T1s_storage}) as a function of time $t_0$, we can plot the average time $\tau_m(x)$ as a function of the mean initial photon number $\bar{n}$ in Fig. \ref{fig:tchar}. Note that $\tau_m(x)$ differs from $1/\Gamma_m$, which is the time needed to reach a fidelity of 92\% (Signal-to-Noise Ratio = 2) when thresholding for a standard qubit readout without decay. But these two timescales give the same order of magnitude for the measurement time.

The increase of $\tau_m$ as a function of the average number of photons can be understood as the fact that the effect of the dissipation increases with the number of photons, so the cavity is increasingly more likely to lose a photon during the measurement time, which increases the time needed by the estimator to converge, as it has to rebuild confidence when a photon is lost. 

\section{Dephasing rate}

\subsection{Measuring the dephasing rate}

To extract the dephasing rate of the cavity in Fig.~4b, we use the technique of Ref.~\cite{Essig2021}. 

\begin{figure}
\centering
\includegraphics[width=\linewidth]{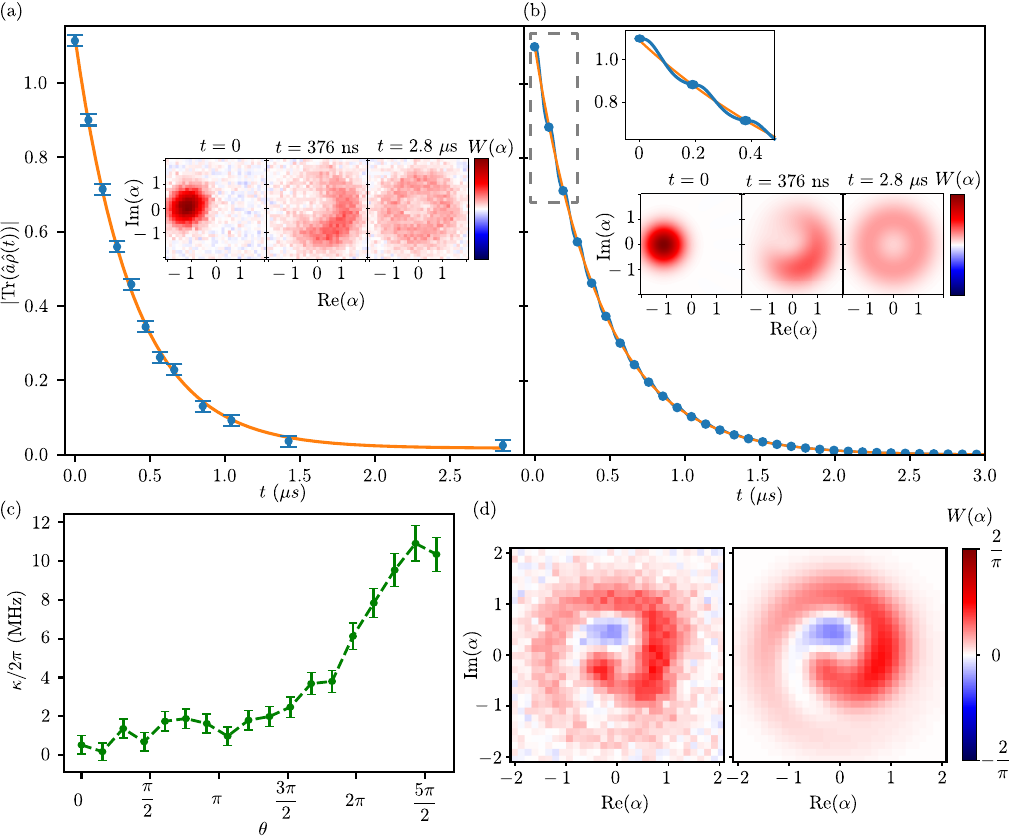}
\caption{(a) Blue dots: measured $|\Tr{\ahat\rhohat(t)}|$ as a function of time for a comb amplitude leading to a qubit kick angle $\theta = \pi$. Solid orange line: fitted exponential decay at a rate $\gamma_\phi$. Inset: timeframes of the measured Wigner functions. (b) Blue line and dots: simulated $|\Tr{\ahat\rhohat(t)}|$ as a function of time for $\theta = \pi$ using Eq.~(\ref{Linblad}). Solid orange line: fitted exponential decay from which the predicted dephasing rate is extracted. Inset: timeframes of the simulated Wigner functions. A zoom on the earliest times shows the dephasing dynamics during a period of the comb drive. (c) Decay rate $\kappa$ of the cavity extracted from the decay of the mean photon number Eq.~(\ref{eq:nbardeWigner}). (d) Measured (left) and simulated (right) Wigner functions at time $t = 15.2~\mathrm{\mu s}$ without driving the qubit. The simulation is performed using Eq.~(\ref{Linblad_cav}) where the self-Kerr rate is $\chi_{cc}/2\pi = 10 \pm1$ kHz, and pure dephasing rate is $\frac{1}{T_{c, \phi}} = (36\pm 6~\mu\mathrm{s})^{-1}$.}
\label{fig:dephasing_exp_simu}
\end{figure}

Performing Wigner tomography via the tomography qubit, we extract $|\langle\hat{a}(t)\rangle|$ after preparing a coherent state $\langle\hat{a}(0)\rangle=-1.11$ in the cavity. In order to cover the full range of dephasing rates, we perform the tomography for multiples of $\pi/\chi \simeq 95$~ns (period of the comb), with an uneven spacing to cover fast and slow dynamics alike. An example of such dynamics with an exponentially decaying amplitude is shown Fig.~\ref{fig:dephasing_exp_simu}a. We use an exponential fit to extract the dephasing rate $\gamma_\phi$. Note that, at small qubit driving amplitude, the decay of $|\langle \hat{a}\rangle|$ is driven by the self-Kerr effect leading to a non-exponential behavior. Fig.~\ref{fig:dephasing_exp_simu}d  shows the measured Wigner function without drive, $15.2~\mathrm{\mu s}$ after the displacement.

If the measurement was QND, we could obtain the measurement induced dephasing rate by subtracting the measured dephasing rate at zero drive $1/T_{c,\phi}$ to the dephasing rate $\gamma_\phi$ at finite drive. However, as explained below, we observe a yet unexplained increase in the cavity decay rate for large qubit drive that forbids such a simple approach. We therefore proceed in two steps.

First, to obtain pure dephasing rates $\Gamma_\phi(\theta)$, we subtract the contribution related to the decay rate $\kappa$ of the cavity to the measured $\gamma$. This decay rate is determined using the tomography at various times. For each measured Wigner function $W_{\rhohat(t)} (\alpha)$, we compute the mean photon number
\begin{equation}
    \overline{n}(t) = \Tr{\adag \ahat \rhohat(t)} \simeq \sum_{k=0}^{7} k \mathbb{P}_k(\rhohat(t)),\label{eq:nbardeWigner}
\end{equation}
where $\mathbb{P}_k(\rhohat)=\int W_{\rhohat} (\alpha) W_{\ketbra{k}}(\alpha)\mathrm{d}^{2}\alpha$ with $W_{\ketbra{k}}(\alpha)$ the Wigner map of $\ketbra{k}$ (see (B4) in~\cite{Essig2021}).
The resulting fitted decay rate $\kappa$ is shown in Fig.~\ref{fig:dephasing_exp_simu}c as a function of the qubit kick angle $\theta$.

Second, in order to extract the measurement induced dephasing rate $\Gamma_d$ plotted in Fig.~4b, we subtract the measured pure dephasing rate without qubit driving $1/T_{c,\phi} = (36\pm 6~\mu\mathrm{s})^{-1}$ to the pure dephasing rate $\Gamma_\phi(\theta)$.

For clarity, the final expression for $\Gamma_d(\theta)$ reads
\begin{equation}
    \Gamma_d(\theta) =\gamma(\theta)-\kappa/2-1/T_{c,\phi}.
\end{equation}

\subsection{Simulation of the dephasing rates}
To predict the expected dephasing rate, we use the Lindblad equation of the qubit-cavity coupled system in the interaction picture
\begin{equation}\label{Linblad}
    \frac{\mathrm{d}\rhohat(t)}{\mathrm{d}t} = -\frac{i}{\hbar}[\hat{H}_\mathrm{int} + \hat{H}_\mathrm{d}, \rhohat] + \frac{1}{T_{q}}\DD_{\sigm} (\rhohat(t)),
\end{equation}
with 
\begin{equation}\label{Hamiltonian qubit-cavity}
   \hat{H}_\mathrm{int}  = -\hbar \frac{\chi}{2} (\mathds{1}+\sigz)\adag\ahat,
\end{equation}
and $\hat{H}_d$ defined in Eq.~(1).

Here, we neglect any other dephasing source, as well as the decay rate of the cavity. We use this simulation to compute $\Tr{\ahat\rhohat(t)}$. Taking only one point per period of the comb. It fits well with an exponential decay (see Fig.~\ref{fig:dephasing_exp_simu}b).

The dephasing rate at zero drive $1/T_{c,\phi}$ can be precisely determined by fitting the dynamics of the cavity state when starting in a coherent state. The Linblad Master equation we use takes into account the self-Kerr effect as well as pure dephasing without drive: 
\begin{equation}\label{Linblad_cav}
    \frac{\mathrm{d}\rhohat(t)}{\mathrm{d}t} = -\frac{i}{\hbar}[\hat{H}_\mathrm{kerr}, \rhohat] + \frac{2}{T_{c, \phi}}\DD_{\adag \ahat} (\rhohat(t)),
\end{equation}
where 
\begin{equation}\label{Hamiltonian_kerr}
\hat{H}_\mathrm{kerr} = -\hbar \chi_{cc} \ahat^{\dag 2} \ahat^2.
\end{equation}
We find that the measured evolution of the cavity state is well reproduced when $\chi_{cc}/2\pi = 10 \pm 1 $ kHz and $1/T_{c, \phi} = (36\pm 6~\mu\mathrm{s})^{-1}$ (see Fig.~\ref{fig:dephasing_exp_simu}d).

\subsection{Predicted dephasing rate}

We start with the same model as in Sec.~\ref{sec:theoretical_rates} where the qubit is kicked once and has infinite time to decay. We can model the experiment as follows. Irrespective of the state of the cavity, the qubit is prepared in a state $\ket{\psi} = \cos{\theta/2}\ket{0} + e^{i\phi}\sin{\theta/2}\ket{1}$. It emits a state $\ket{\phi}_n$ in the transmission line that depends on the number of photons $n$, which we can write 
\begin{align}
    \ket{\phi}_n = \Big(\cos(\theta/2)\mathds{1} + e^{i\phi}\sin(\theta/2)\sqrt{\frac{1}{T_q}}\int_0^{+\infty}e^{-i(\omega_q-n\chi)t-\frac{t}{2T_q}}\ahat_\mathrm{out}^\dagger(t)\mathrm{d}t\Big)\ket{\mathrm{vac}}
\end{align}
where $\ket{\mathrm{vac}}$ is the vacuum state of the outgoing transmission line.
In the simple case where $n= 0$ or $n=1$, the cavity can be treated as a qubit as in Ref.~\cite{Clerk2010IntroductionAmplification}. Each period $\tau = \frac{\pi}{\chi}$ of the train of pulses leads to a decay of the off-diagonal matrix elements of the cavity density matrix by a factor $|\tensor[_0]{\braket{\phi}{\phi}}{_1}|$. The dephasing rate $\Gamma_d$ (black solid line in Fig.~4b) is thus given by
$$e^{-\Gamma_d \tau} = |\tensor[_0]{\braket{\phi}{\phi}}{_1}|$$
where
$$ \tensor[_0]{\braket{\phi}{\phi}}{_1} =\cos^2(\theta/2) +\frac{ \sin^2{\theta/2}}{1 -i\chi T_q }.$$

\section{Deviation to Quantum Nondemolition}
\begin{figure*}[b] 
\centering
\includegraphics[width=\linewidth*1/3]{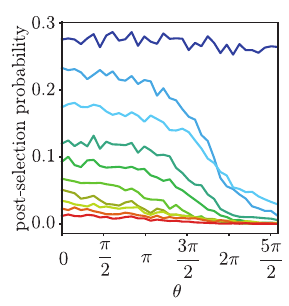}
\caption{Probability of measuring $\mathrm{Re}(r)$ below threshold during the post-selection pulse of the sequence in Fig.~\ref{fig:selection_sequence}a as a function of the qubit rotation angle $\theta$ associated to the comb. Each line corresponds to a different selected photon number $n$ from $0$ (dark blue) to $9$ (red). The selection thresholds are the same as in Table~\ref{table_yield}}.
\label{fig:postselectionyield}
\end{figure*}

\begin{figure}[!]
\centering
\includegraphics[width=\linewidth]{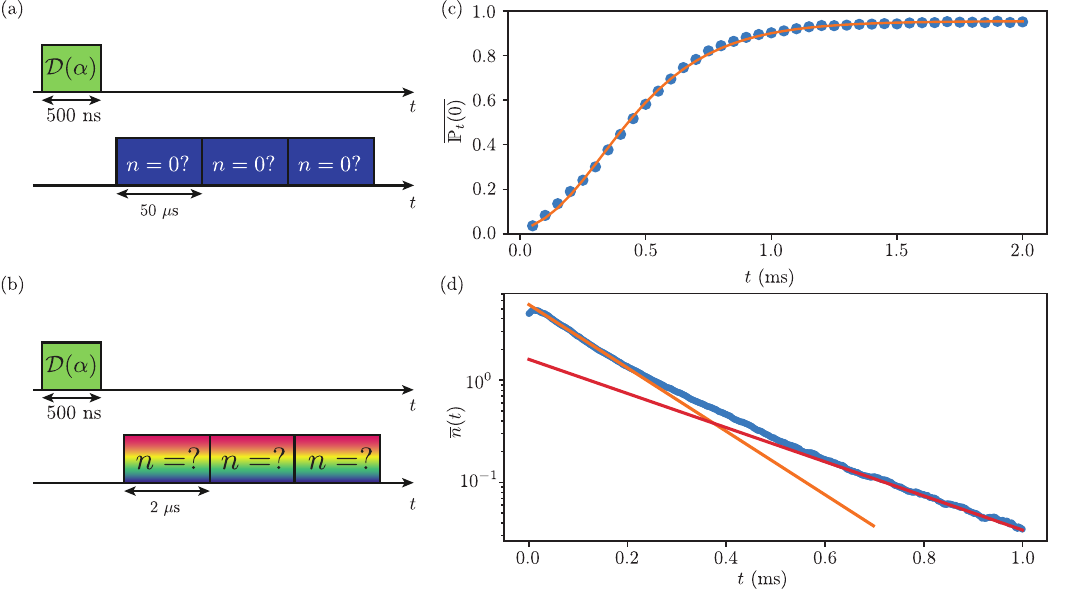}
\caption{(a) Pulse sequence tracking the occupation of the vacuum state. (b) Pulse sequence tracking the number of photons. (c)  Blue dots: probability $\overline{\mathds{P}_t (n=0)}$ that the cavity is in vacuum as a function of time. The measurement is obtained by single-tone readout of the number of photons after a strong ($|\alpha|^2>20$) displacement of the memory. The fit (orange) to Eq.~(\ref{eq:pde0oft}) gives $T_c =230~\mu$s. The absence of a plateau at early times despite the large displacement could be attributed to an increasing decay rate for higher number of photons. (d) Blue dots: reconstructed average number of photons using 1100 records of the tracked photon number. Exponential fits with $T_c = 130~\mu$s (orange) and $T_c = 250~\mu$s (red) are superimposed.}
\label{fig:T1s_storage}
\end{figure}

In order to better characterize how the driving amplitude of the qubit affects the number of photons in the cavity (deviation to QNDness of the photon number tracking), we perform several experiments. The drive amplitude is given either as a Rabi frequency $\Omega$ for single tone driving or an angle of rotation $\theta$ during a qubit kick.

A direct signature of non-QNDness can be seen in Fig.~\ref{fig:postselectionyield}. It shows the yield of the post-selection for each $n$ (each color) as a function of the rotation angle $\theta$. The yield decreases with driving amplitude, showing that driving the qubit changes the transition rates between Fock states. The variation is more pronounced for larger driving amplitudes ($\theta>\pi$). As a matter of fact, the curves for $n= 1$ and $n=2$ cross, which indicates a decay rate that changes with $n$. 

In Fig.~\ref{fig:dephasing_exp_simu}c, we show the observed cavity decay rate $\kappa$ as a function of the angle $\theta$. It is obtained by computing the number of photons as a function of time from the Wigner tomography. A small increase of $\kappa$ seems to occur as $\theta$ grows but still an order of magnitude smaller than the measurement rate for $\theta=\pi/2$. However a sharp rise in $\kappa$ occurs for $\theta\approx 3\pi/2$.

\section{Dependence of the cavity lifetime on photon number}

We also performed a measurement of the average probability of having 0 photons  $\overline{\mathds{P}_t(n=0)}$  as a function of time (Fig.~\ref{fig:T1s_storage}a,c). This is obtained with a single frequency tone on the qubit at $\Omega=5.4~\textrm{MHz}$ after a cavity displacement. Fitting it with the decay of a coherent state 
\begin{equation}
\overline{\mathds{P}_t (n=0)} = |\langle\sqrt{n_0} e^{-t/T_{c}}|0\rangle|^2= e^{-n_0 e^{-t/T_{c}}}\label{eq:pde0oft}
\end{equation} extracts the decay time $T_c$ of one photon in the cavity.

We do not observe a plateau at initial times (small effective $n_0$ in Eq.~(\ref{eq:pde0oft})) despite the large variation of displacement amplitudes we tested. We believe that this could be due to a nonlinear decay rate of the cavity owing the presence of the transmon and its Purcell filter. 

Using the quantum trajectories obtained with the comb drive, we can also reconstruct the average number of photons as a function of time. It does not  decay exponentially. This non exponential cavity decay was also observed using the tomography qubit on a different run. As a side remark, the slight increase of the estimated mean number of photons at small times is well understood. Indeed the comb signal is only analyzed within the range from 0 to 9 photons so that a wrong prior estimates the average number of photons at 4.5 initially. As the number of photons decreases and enters the detecting window $[\![0, 9]\!]$, the update tends to increase the average number of photons at short times.

\begin{figure}[!]
\centering
\includegraphics[width=\linewidth]{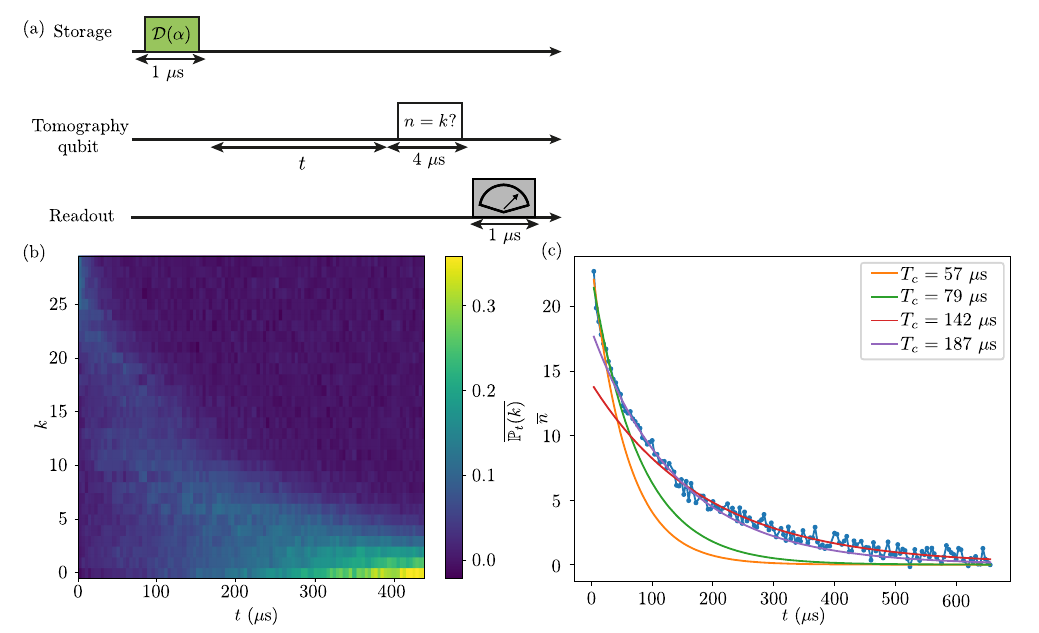}
\caption{(a) Pulse sequence for the measurement of the probability $\overline{\mathds{P}_t(k)}$ that there are $k$ photons in the resonator. (b) Measured $\overline{\mathds{P}_t(k)}$ as a function of time $t$ and photon number $k$ after a displacement pulse of about 25 mean photons.(c) Measured decay of the average photon number $\overline{n}=\sum_{k=0}^{30} k\overline{\mathds{P}_t(k)}$ as a function of time.}
\label{fig:T1s_storage_YN}
\end{figure}

In a previous cool down of the same device, we measured the decay of the cavity for a wide range of numbers of photons using the tomography qubit. The procedure consists in first displacing  the cavity to a coherent state with about 25 average photons. The tomography qubit is then driven with a selective $\pi$ pulse (at a frequency $\omega_t-k\chi_{ct}$, see Table \ref{circuits parameters}), conditioned on the number of photons being equal to a certain $k$. The state of the tomography qubit is then read out using its readout resonator. From this, we can reconstruct the average occupation of each Fock state $\overline{\mathds{P}_t(k)}$ in the cavity at any time $t$ after the displacement (Fig.~\ref{fig:T1s_storage_YN}b). The behaviour of $\overline{\mathds{P}_t(k)}$ is  non exponential for high number of photons, and seems to drop faster for more than 10 photons.
Looking at the average number of photons on Fig.~\ref{fig:T1s_storage_YN}b, we fit the curve portions with several exponential functions, and see that the decay rate of the cavity increases with the average number of photons.

We do not have a simple explanation for this behaviour. We believe that higher order conversion processes between the cavity and the multiplexing qubit could be involved in these observations. As a matter of fact, the frequency of the qubit is higher than that of the cavity, which could more easily trigger the effects described in \cite{Khezri2022}. However, it would be in a very unusual regime, as the cavity has a long decay time and the qubit a low one in our work.

